\documentclass[twocolumn]{aastex631} 
\newcommand{\Lsun}{$\rm L_{\odot}$}
\newcommand{\Msun}{$\rm M_{\odot}$}

\newcommand{\Mdot}{$\dot{M}$}

\newcommand{\Pdot}{$\dot{P}$}
\newcommand{\Av}{$A_V$}

\newcommand{\mic}{$\rm \mu$m}

\usepackage{amsmath}

\shorttitle{Jet Precession in NGC 2071 IR}
\shortauthors{Rubinstein et al.}

\begin{document}

\title{HOPS 361-C's Jet Decelerating and Precessing Through NGC 2071 IR}

\correspondingauthor{Adam E. Rubinstein}
\email{arubinst@ur.rochester.edu}


\author[0000-0001-8790-9484]{Adam E. Rubinstein}
\affiliation{Department of Physics and Astronomy, University of Rochester, Rochester, NY 14627-0171, USA}

\author[0000-0003-3682-854X]{Nicole Karnath}
\affiliation{Space Science Institute, 4765 Walnut St, Suite B
Boulder, CO 80301, USA} 
\affiliation{Universities Space Research Association, MS 232-11, NASA Ames Research Center, Moffett Field, CA 94035, USA} 
\affiliation{Center for Astrophysics Harvard \& Smithsonian, Cambridge, MA 02138, USA}

\author[0000-0003-1280-2054]{Alice C. Quillen}
\affiliation{Department of Physics and Astronomy, University of Rochester, Rochester, NY 14627-0171, USA}

\author[0000-0002-6136-5578]{Samuel Federman}
\affiliation{Ritter Astrophysical Research Center, Department of Physics and Astronomy, University of Toledo, Toledo, OH 43606, USA}

\author[0000-0003-1665-5709]{Joel D. Green}
\affiliation{Space Telescope Science Institute, 3700 San Martin Drive, Baltimore, MD 21218, USA}

\author[0000-0003-4195-1032]{Edward T. Chambers}
\affiliation{Universities Space Research Association, MS 232-11, NASA Ames Research Center, Moffett Field, CA 94035, USA} 

\author[0000-0001-8302-0530]{Dan M. Watson}
\affiliation{Department of Physics and Astronomy, University of Rochester, Rochester, NY 14627-0171, USA}

\author[0000-0001-7629-3573]{S. Thomas Megeath}
\affiliation{Ritter Astrophysical Research Center, Department of Physics and Astronomy, University of Toledo, Toledo, OH 43606, USA}

\begin{abstract}

We present a two-epoch Hubble Space Telescope (HST) study of NGC 2071 IR highlighting HOPS 361-C, a protostar producing an arced 0.2 parsec-scale jet. Proper motions for the brightest knots decrease from 350 to 100 km/s with increasing distance from the source. The [Fe II] and Pa$\beta$ emission line intensity ratio gives a velocity jump through each knot of 40--50 km/s. A new [O I] 63 \mic\ spectrum, taken with the German REciever for Astronomy at Terahertz frequencies (GREAT) instrument aboard Stratospheric Observatory for Infrared Astronomy (SOFIA), shows a low line-of-sight velocity indicative of high jet inclination. Proper motions and jump velocities then estimate 3D flow speed for knots. Subsequently, we model knot positions and speeds with a precessing jet that decelerates. Measurements are matched with a precession period of 1,000--3,000 years and half opening angle of $15^\circ$. The [Fe II] 1.26-to-1.64 \mic\ line intensity ratio determines visual extinction to each knot from 5--30 mag. Relative to $\sim$14 mag of extinction through the cloud from $\rm{C}^{18}$O emission maps, the jet is embedded at a 1/5 to 4/5 fractional cloud depth. Our model suggests the jet is dissipated over a 0.2 pc arc. This short distance may result from the jet sweeping through a wide angle, allowing the cloud time to fill cavities opened by the jet. Precessing jets contrast with nearly unidirectional protostellar jets that puncture host clouds and can propagate significantly further.

\end{abstract}


\section{Introduction} 
NGC 2071 is a region of embedded star formation in the Orion B molecular cloud complex and is rich in intermediate-mass protostars and protostellar jets. The setting is ideal for investigation of interactions between outflows and the host molecular cloud \citep{Bally_1982, Wootten_1984,Bally_2016,Lee_2020}. 
%
At IR wavelengths the northern region is dubbed NGC 2071 IR. It contains bright, protostars like HOPS 361-C (IRS 3) and HOPS 361-A (IRS 1) \citep{Walther_2019}, both of which feature prominent jets traced in molecular hydrogen emission  \citep{Eisloeffel_2000}. HOPS 361-A, one of the brightest sources in NGC 2071 IR, has a distance of 430.4 pc using Gaia observations \citep{Tobin_2020}. 
The region and its protostars were recently studied at high-angular resolution with the Atacama Large Millimeter/submillimeter Array (ALMA) and the Very Large Array (VLA) by \citet{Cheng_2022}.


HOPS 361-C is the proposed driving source for the largest NE-SW outflow bright in molecular hydrogen. The outflow features large $\sim 3'$ arcs on both sides of the HOPS 361 region \citep{Eisloeffel_2000}. HOPS 361-C is a Class 0 protostar \citep{Tobin_2020}, although it is borderline with Class I according to its bolometric temperature of 69 K \citep{Furlan_2016}. A Keplerian molecular disk is resolved with ALMA in SO $6_5$ -- $5_4$, with an outer disk radius of $\sim$150 au and position angle of $\sim 130^\circ$ approximately perpendicular to a high-velocity CO jet \citep{Cheng_2022}. 

Misaligned segments and wiggles in the outflow tracer, CO J = 2--1, led \citet{Cheng_2022} to propose that the molecular jet is produced by a young stellar object with a precessing disk. A jet that precesses due to tidal interactions in a binary system may experience wide variations in direction. These variations occur when the jet axis is tilted or misaligned with respect to the direction normal to the binary's orbital plane \citep{Terquem_1999,Masciadri_2002}.  

Variations in jet direction can also affect stellar feedback and ensuing star formation. 
If the jet remains at a fixed angle, then a single cavity is cleared within a molecular cloud \citep{Cunningham_2009b}.  
In contrast, arced jets can affect a larger volume of the undisturbed cloud
\citep{Hartigan_2001} or be dissipated locally \citep{Fendt_2022}. 
By comparing properties of wide, arced jets with those that exhibit small opening angles, we can constrain how the jet propagates into the ambient molecular cloud, an important step toward quantifying the reach of feedback. 

Bright gaseous knots trace shocks in protostellar jets and outflows, and their positions along a jet can wiggle (e.g., \citealt{Raga_2002}). These wiggles have been attributed to variations in jet velocity caused by orbital motion within a binary system or by variations in jet orientation due to precession \citep{Masciadri_2002,Anglada_2007,Erkal_2021}. If both a jet and counter-jet are observed, the two scenarios can be differentiated via examination of the symmetry pattern of the jet trajectory. Mirror-symmetry is expected in the case of binary orbital motion; an S-shaped morphology is expected in the case of precession. These two types of models can be applied to the jet and knot positions, directly relating periodicity in the knot positions to the binary period or precession period  \citep[e.g.][]{Raga_2002}. Most prior work focuses on molecular jets with small wiggles and opening angles (i.e. the angular width of the jet) of $<$10$^\circ$; these are usually interpreted in terms of binary motion \citep[][]{Gueth_1996, Zinnecker_1998, Woitas_2002, Chandler_2005, Sahai_2005, Ybarra_2006, Seale_2008, Phillips_2009, Lee_2010, Whelan_2010, AgraAmboage_2011, NoriegaCrespo_2011, Estalella_2012, FernandezLopez_2013, Velazquez_2014, Kwon_2015, Lee_2015, Beltran_2016, Moraghan_2016, Choi_2017, Louvet_2018, NoriegaCrespo_2019, Hara_2021, Jhan_2021, Murphy_2021, Massi_2022, Massi_2023}.

Some jets exhibit a visible S-shape but have a narrow opening angle of 8$^\circ$, like the L1157 molecular outflow and jet \citep{Podio_2016}. 
The outflow from HOD07 1 in Monoceros R2 \citep{Hodapp_2007}, from IRAS 4A2 in NGC 1333 \citep{Hodapp_2005, Jorgensen_2006, Chuang_2021}, the jet in Barnard 1 denoted B1c \citep{Matthews_2006}, and the collimated outflows surveyed in the Cores to Disks Spitzer Legacy program \citep{Seale_2008} all show an S-shaped morphology and are likely to have wider opening angles. S-shaped jets and jets that appear to have a wide opening angle are often interpreted in terms of only precession of the binary orbit \citep{Hodapp_2005, Jorgensen_2006,Matthews_2006,Podio_2016, Chuang_2021, Lee_2020}. However, \citet{Cunningham_2009} proposed that the HW2 protostellar source in the Cepheus A region precessed due to capture of a binary partner, but this result cannot entirely rule out the binary precessing \citep{Ferrero_2022}.  Variations in jet direction can also be due to  asymmetric infall from the envelope or tidal torque from the envelope resulting in disk precession \citep{Hirano_2019}. The massive protostar IRAS 20126+4104 has a precessing jet with the widest known opening angle of 40$^\circ$, but CO imaging did not distinguish between possible interpretations \citep{Shepherd_2000}. 

Sub-arcsecond resolution and multi-epoch observations are required to study the morphology and motion of bright knots along a protostellar jet \citep{Bally_1982,Reiter_2017, Hartigan_2019, Lopez_2022}. 
Near-infrared (NIR) emission lines can be used to detect these knots in heavily extinguished regions \citep[e.g.][]{Erkal_2021}.
Shocks throughout a jet consist of sharp, discontinuous jumps in density and velocity, called Jump or J-type, and Continuous versions or C-type.
To separate emission from dust and from shocks, forbidden and atomic spectral lines, like [Fe II] and Pa$\beta$ respectively, are used to determine extinction and to study hot, ionized, and shocked gas in protostellar jets \citep[e.g.][]{Erkal_2021, Reiter_2017}. 

We focus on HOPS 361-C, which has a clear and distinct string of knots tracing an arced jet with a wide opening angle. In Section \ref{obs} we present new Hubble Space Telescope (HST) observations of NGC 2071 IR in [Fe II] and Pa$\beta$ and NIR HST images from an earlier epoch. Knots of emission are detected, presumed from J-type shocks associated with the jet. We also present a new [O I] spectrum using the German REciever for Astronomy at Terahertz frequencies (GREAT) instrument on the Stratospheric Observatory for Infrared Astronomy (SOFIA), which measures the radial velocity near the base of the jet, and constrains the orientation of the outflow relative to the plane of the sky. In Section \ref{analysis_promot} we use the multi-epoch data to measure knot proper motions. In Section \ref{analysis_shocks} we discuss constraints on shock models and use [Fe II] line ratios to calculate the extinction to each knot in the jet. In Section \ref{discussion} we discuss models that can account for the speed and morphology of the jet originating from HOPS 361-C. Shock speeds from our spectral line images combine with tangential speeds from proper motions to give the jet’s momentum and kinetic energy injection rate into the cloud. With our model of HOPS 361-C’s jet, we estimate the rate that jet material is decelerated by interacting with the molecular cloud. 
    \label{intro}

\section{Observations and Data Reduction}  
    \label{obs}
    
    \subsection{HST WFC3/IR}
    NGC 2071 was observed twice with HST's Wide Field Camera 3/Infrared (WFC3/IR).  
The first epoch is centered on NGC 2071 IR and the protostellar sources HOPS 361(A--F). These images were taken on Mar 8, 2010 from HST General Observer (GO) proposal 11548 \citep{Kounkel_2016, Habel_2021}. 
The second epoch of images were taken between Sept 16 and Nov 3, 2021 as part of a joint HST/SOFIA GO proposal 16493. 
That program observed three fields to cover the entire HOPS 361-C jet, the HOPS 361-A outflow cavity, and the region surrounding these features.

The image frames were acquired with HST/WFC3 in star tracker-guided mode and using a standard 4-point dither. WFC3's field of view is 136" x 123" at NIR wavelengths. The images were captured using the following narrowband, NIR filters (and relevant spectral features): F126N (1.26 \mic~[Fe II]), F128N (1.28 \mic~Pa$\beta$), F130N (1.30 \mic~continuum), F164N (1.64 \mic~[Fe II]), F167N (1.67 \mic~continuum). Images are standard data products outputted by the {\tt calwf3} pipeline, retrieved from the Barbara A. Mikulski Archive for Space Telescopes (MAST) with suffix "FLT", and have details in Table \ref{tab:obs}.

\startlongtable
\begin{deluxetable*}{l c c c c c}
    \tabletypesize{\footnotesize}
    \tablecaption{Observing Details \label{data}}
    \tablehead{ 
        \colhead{Target} & \colhead{Observation ID} & \colhead{Date} & \colhead{Filter} & \colhead{Spectral Line and Wavelength} & \colhead{Exposure Time (sec)} 
    }
    \startdata
        HOPS 369 & IB0L9X010 & 2010 Mar 8  & F160N & Continuum 1.60 \mic & 2496 \\
        HOPS361-center & IEJ707010  & 2021 Nov 1  & F126N & [Fe II] 1.257 \mic & 1797.7 \\
        HOPS361-center & IEJ708010  & 2021 Nov 1 & F128N  & H I Pa$\beta$ 1.282 \mic & 1797.7 \\
        HOPS361-center & IEJ707020,IEJ708020  & 2021 Nov 1  & F130N & Continuum 1.30 \mic & 597.7,597.7 \\
        HOPS361-center & IEJ701010  & 2021 Sept 16  & F164N & [Fe II] 1.644 \mic & 2396.9 \\
        HOPS361-center & IEJ702010  & 2021 Sept 16 & F167N & Continuum 1.67 \mic & 2396.9\\
        HOPS361-SW & IEJ705010 & 2021 Sept 30 & F164N & [Fe II] 1.644 \mic &  2396.9 \\
        HOPS361-SW & IEJ706010 & 2021 Nov 1 & F167N & Continuum 1.67 \mic &  2396.9  \\
        HOPS361-SW & IEJ711010 & 2021 Nov 3 & F126N & [Fe II] 1.257 \mic & 1797.7 \\
        HOPS361-SW & IEJ711020,IEJ712020 & 2021 Nov 3 & F130W & Continuum 1.30 \mic & 597.7,597.7 	\\ 
        HOPS361-SW & IEJ712010 & 2021 Nov 3 & F128N & H I Pa$\beta$ 1.282 \mic & 1797.7 \\
        HOPS361-NE & IEJ703010 & 2021 Sept 27 & F164N & [Fe II] 1.644 \mic & 2396.9\\
        HOPS361-NE & IEJ704010 & 2021 Sept 29 & F167N & Continuum 1.67 \mic & 2396.9\\
        HOPS361-NE & IEJ709010 & 2021 Nov 2 & F126N & [Fe II] 1.257 \mic & 1797.7	 \\ 
        HOPS361-NE & IEJ709020,IEJ710020 & 2021 Nov 2 & F130N & Continuum 1.30 \mic & 597.7,597.7\\
        HOPS361-NE & IEJ710010 & 2021 Nov 2 & F128N & H I Pa$\beta$ 1.282 \mic & 1797.7 \\
    \enddata
    \tablecomments{Target and Observation IDs taken from MAST ordered by date of observation. Target centers (J2000 RA, Dec) were
    HOPS369: (05h47m01.606s, +00d17m58.88s), 
    HOPS361-Center: (05h46m58.943s, +00d20m41.12s),
    HOPS361-SW: (05h46m58.943s, +00d20m41.12s),
    HOPS361-NE: (05h47m09.900s, +00d23m46.72s).\label{tab:obs}}
\end{deluxetable*}

We aligned images using the Python {\tt drizzlepac} package with default parameters. Images in all filters were simultaneously aligned using the {\tt TweakReg} routine Version 1.4.7 to give a sub-pixel accuracy. 
We estimate errors in the image shifts to
be  at most 0.05 pixels or $\sim$ 0.006" based on comparing centroids of stars in images at the two different epochs. 

The {\tt Astrodrizzle} algorithm \citep{Hoffmann_2021} Version 3.3.0 was used with default parameters to resample the images to the same pixel size of $0''.12825 \times 0''.12825$ and remove defects (e.g. cosmic rays, bad pixels). The same routine was used to mosaic sets of second epoch narrowband images with the same filter. 
We then used the IPAC software {\it Montage} to shift, regrid, and update the World Coordinate System to the same frame size and orientation. 
This gave a final spatial (angular) resolution by 2D Gaussian FWHM of about 2 pixels (0.2565 arcsec) for epoch 1 and about 3 pixels (0.38475 arcsec) for epoch 2.


To convert e-/s to units of intensity, we used the flux density and bandwidth calibration values in the FITS file headers produced by the HST calibration pipeline (keywords {\tt PHOTFLAM} and {\tt PHOTBW}).
The 
RMS noise in the central frame for all images and within circular areas ranging from 1--15 $\rm arcsec^2$ (used through this work) is 1.55 $\times {10}^{-18}\ {\rm erg}\ {\rm cm}^{-2}\ {\rm s}^{-1}$ per pixel as measured in regions devoid of stars and nebulosity.

\begin{figure*}
    \begin{minipage}[b]{0.5\linewidth}
    \centering
    \includegraphics[width=\textwidth, trim = 0.55in 0.55in 0in 0in,clip]{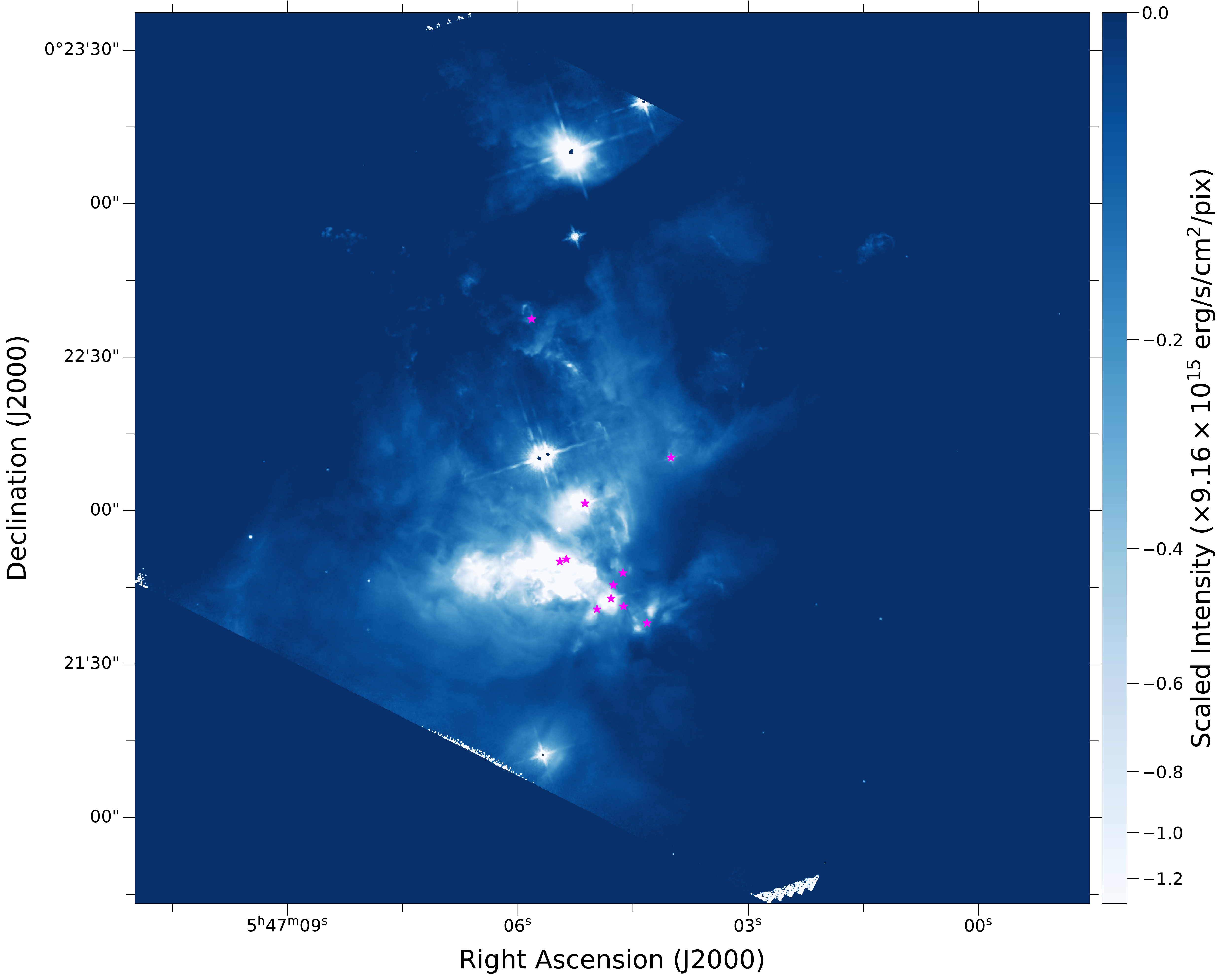}
    \end{minipage}
    \begin{minipage}[b]{0.5\linewidth}
    \centering
    \includegraphics[width=\textwidth, trim = 0.55in 0.55in 0in 0in,clip]{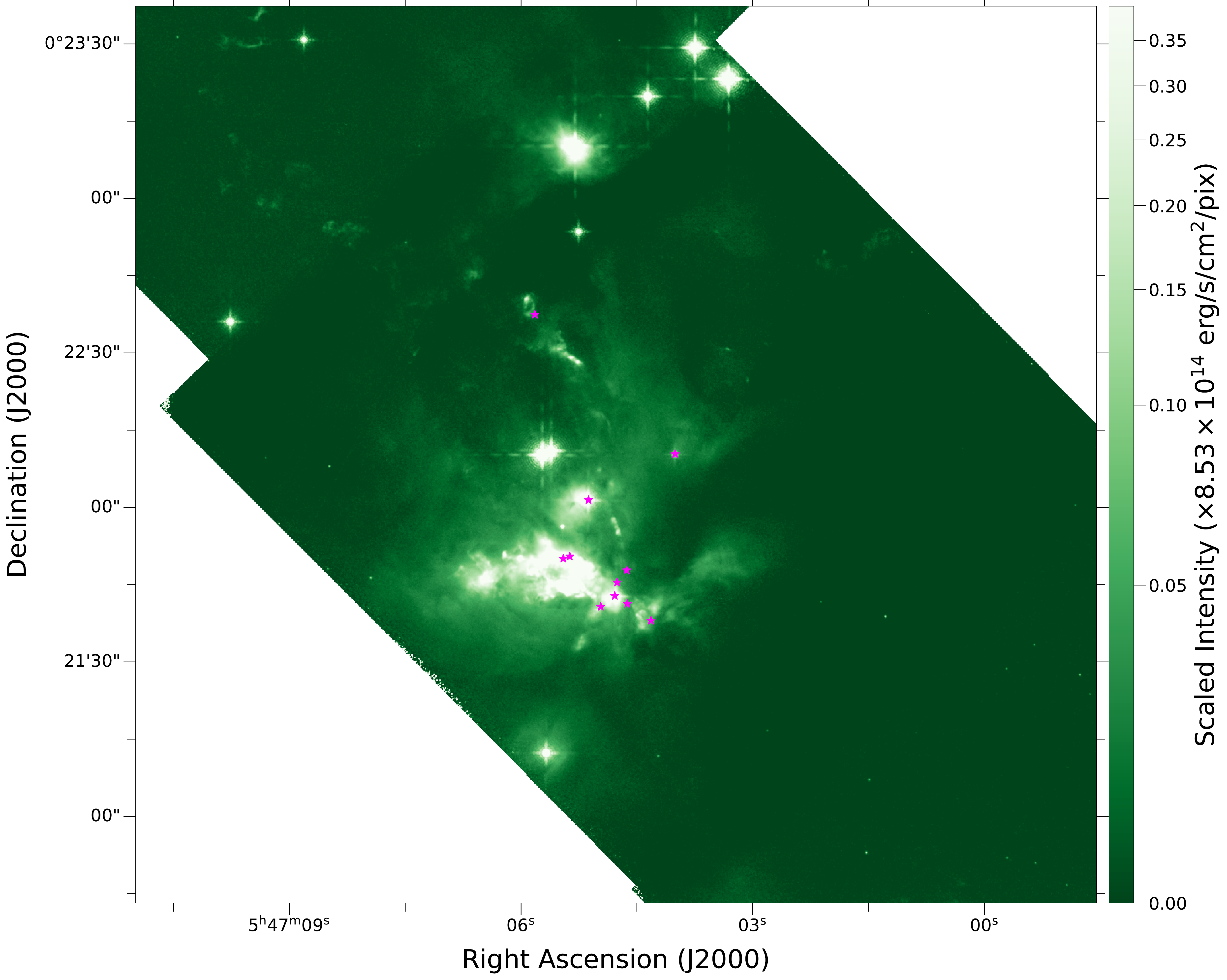}
    \end{minipage}
    
    \begin{minipage}[b]{\linewidth}
    \centering
    \includegraphics[width=\textwidth, trim = 0in 0in 0in 0in,clip]{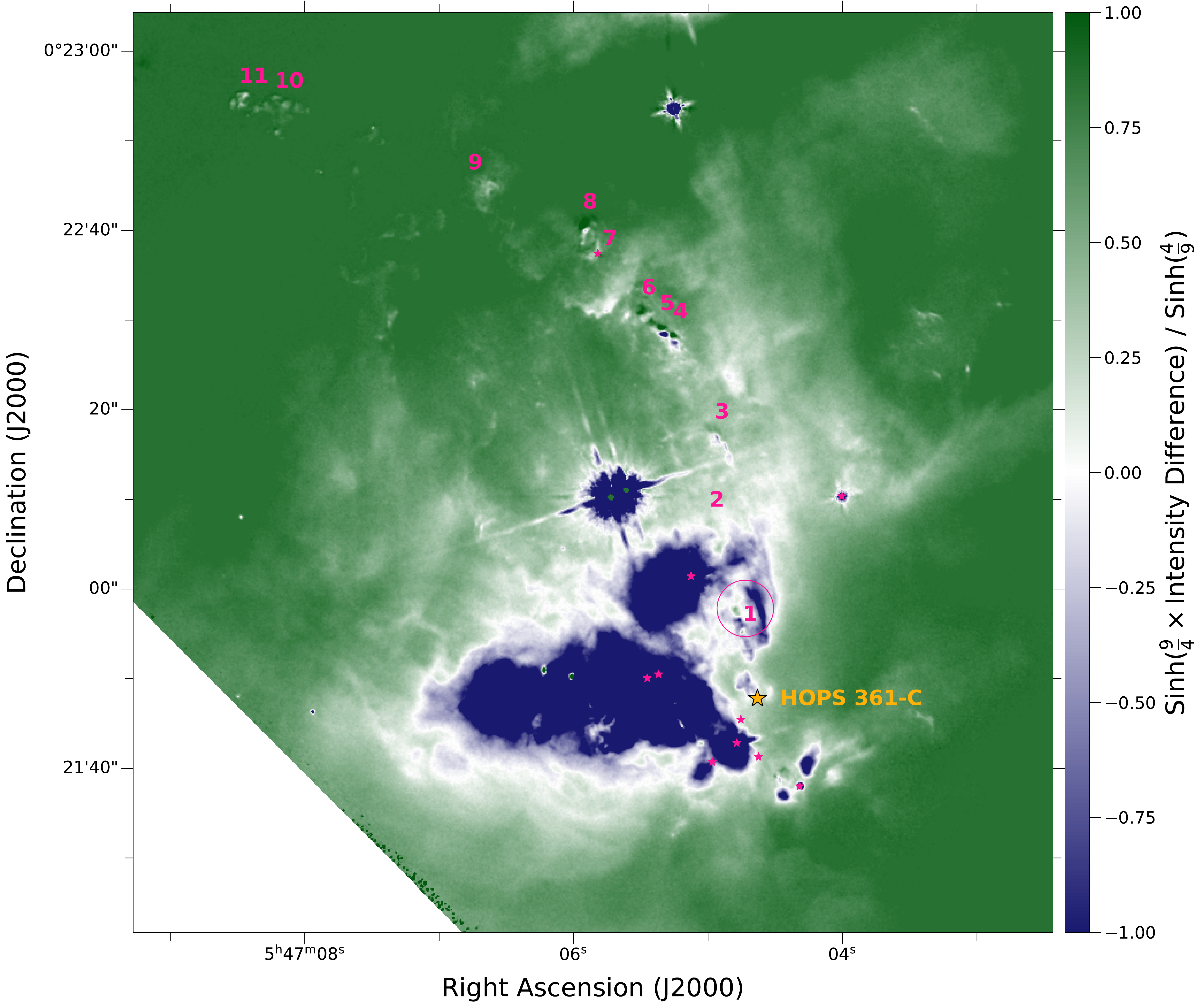}
    \end{minipage}
    
    \caption{Images centered on the HOPS 361 region. 
    \underline{\textit{Top left}}: Epoch 1 F160W image with a Sinh stretch and
     (\underline{\textit{top right}}) epoch 2 synthetic F160W image with an Arcsinh stretch.
    (\underline{\textit{Bottom}}) The cropped F160W difference image with a diverging color bar and Sinh stretch. Blue (negative values) represents epoch 1 and green (positive values) epoch 2; blue transitioning to green marks a white boundary for a knot potentially moving between epochs. 
    HOPS 361-C is shown with the largest, yellow, star-shaped marker, and other nearby protostellar sources are pink. 
    Knots we identify (Table \ref{tab:knot_measures}) are numbered for reference, where 1 is closest and 11 is furthest from HOPS 361-C.
    Our SOFIA 4GREAT beam is shown around knot 1.
    }
    \label{fig:compare_epochs}
\end{figure*}

For the precise co-alignment needed to compute proper motions, the F160W frames were separately reprocessed through with {\tt drizzlepac}, {\tt Astrodrizzle}, and \textit{Montage} as described with the corresponding second epoch F164N and F167N frames covering the closest overlapping area. 
Our images from both epochs centered on HOPS 361 are shown in the top left and top right of Figure \ref{fig:compare_epochs}. HOPS 361-C has an R.A. of 5h47m4.631s and Dec. of +0d21m47.82s according to ALMA data in \citet{Tobin_2020}. This diverges from the coordinates found by \citet{Walther_2019}, which are based on 2MASS K-band images, but we choose to use those from \citet{Tobin_2020} because they were found with higher spatial resolution.

\begin{figure*}
    \centering
    \includegraphics[width=0.5\textwidth, trim = 0.75in 0.75in 0in 0in,clip]{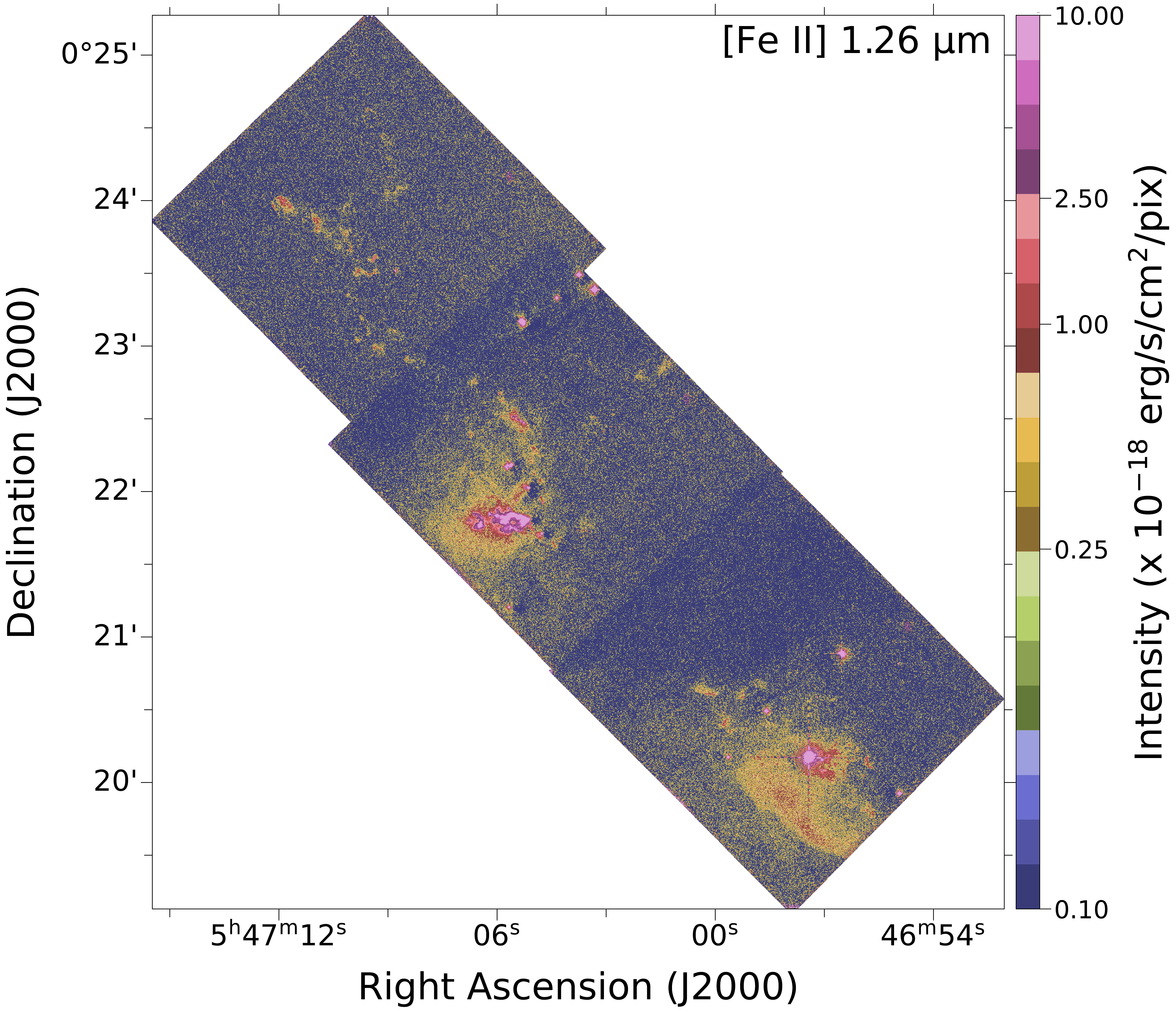}
    \includegraphics[width=0.5\textwidth, trim = 0.75in 0.75in 0in 0in,clip]{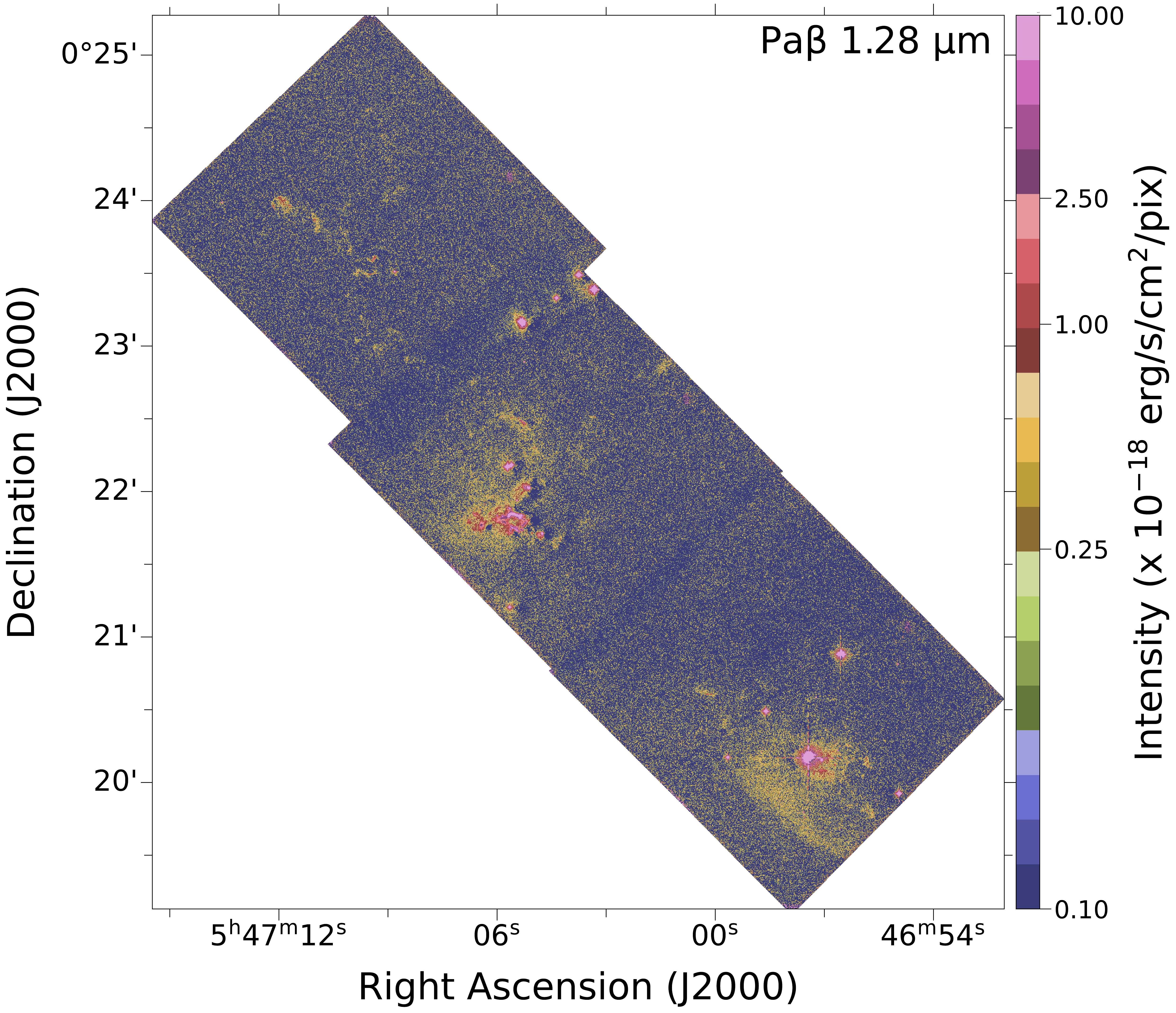}
    \includegraphics[width=0.5\textwidth, trim = 0in 0in 0in 0in,clip]{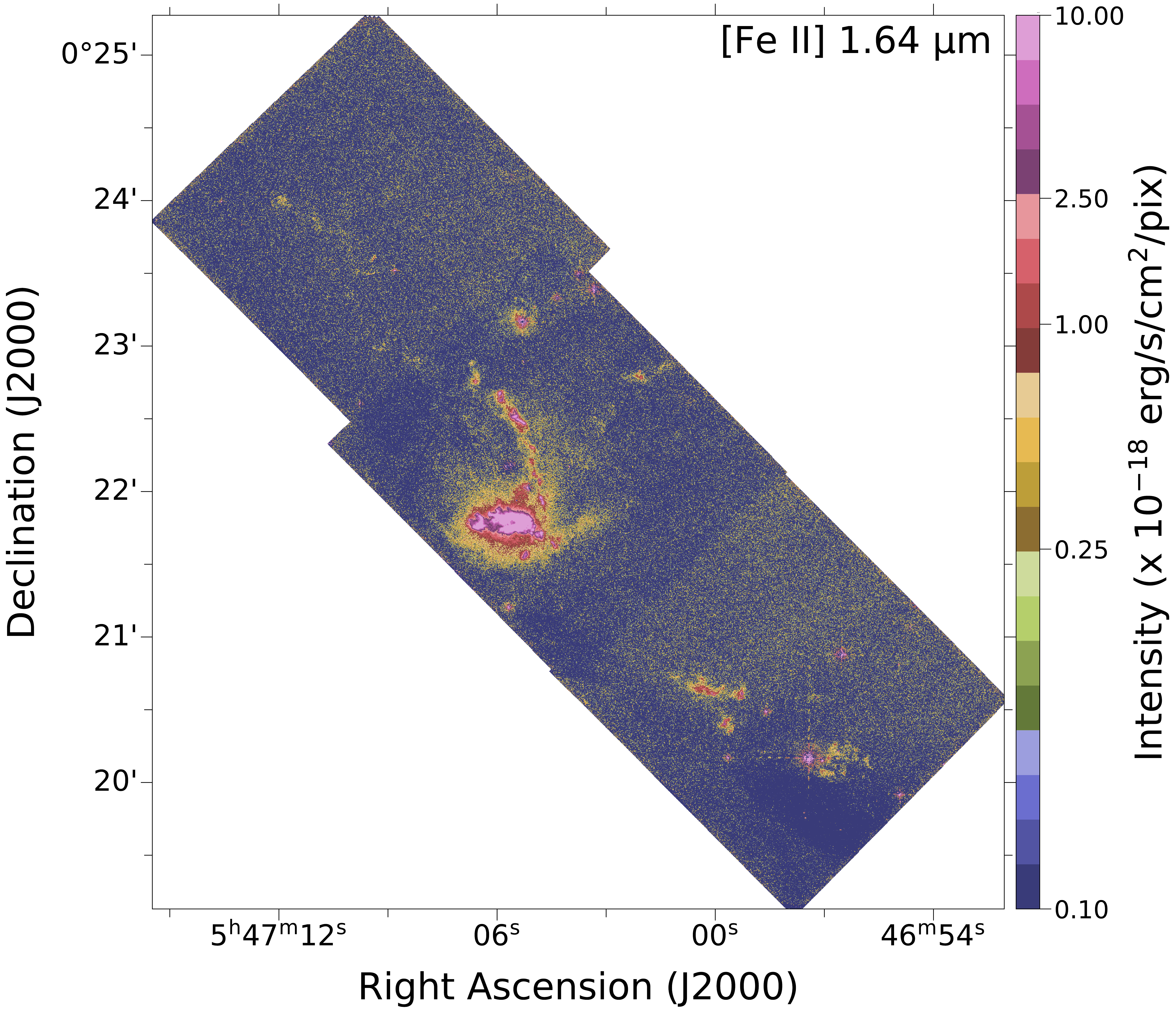}
    \caption{From top to bottom, we show the continuum-subtracted 1.26 \mic~[Fe II], 1.28 \mic~Pa$\beta$, and 1.64 \mic~[Fe II] mosaics co-aligned  for ease of comparison. All colorbars use a natural log stretch. }\label{fig:mosaics_epoch2}
\end{figure*}

For the second epoch images, the closest continuum filter was subtracted from each line image to construct narrowband images in [Fe II] 1.26 \mic~ and 1.64 \mic~ as well as Pa$\beta$~ 1.28 \mic. 
 The appropriate continuum filters were F130N for F126N and F128N, and F167N for F164N.
The continuum-subtracted narrowband images from the second epoch are shown in Figure \ref{fig:mosaics_epoch2}.

    \label{hst_obs}

    \subsection{SOFIA GREAT}
    SOFIA observations taken on Feb 9--10, 2021 use the 4GREAT channel to observe the [O I] 63 $\mu$m line. 
The 4GREAT footprints overlap the HOPS 361 region,
and [O I] emission 
traces the shocked material to give a jet's radial velocity and calibrate shock models, as shown with observations of other emission lines  \citep[e.g.][]{Hartmann_1984, Carr_1993, Hartigan_2000, Graham_2003} and theoretically for jets exhibiting bow shocks \citep[e.g.][]{Hartigan_1987}. The observations have a sensitivity of 0.04~K with 2 km/s bins, resulting in a signal-to-noise ratio (SNR) of $\sim$6. 

\begin{figure*}
        \includegraphics[width=0.6\textwidth, angle=270, trim = 2in 0in 0in 0in,clip]{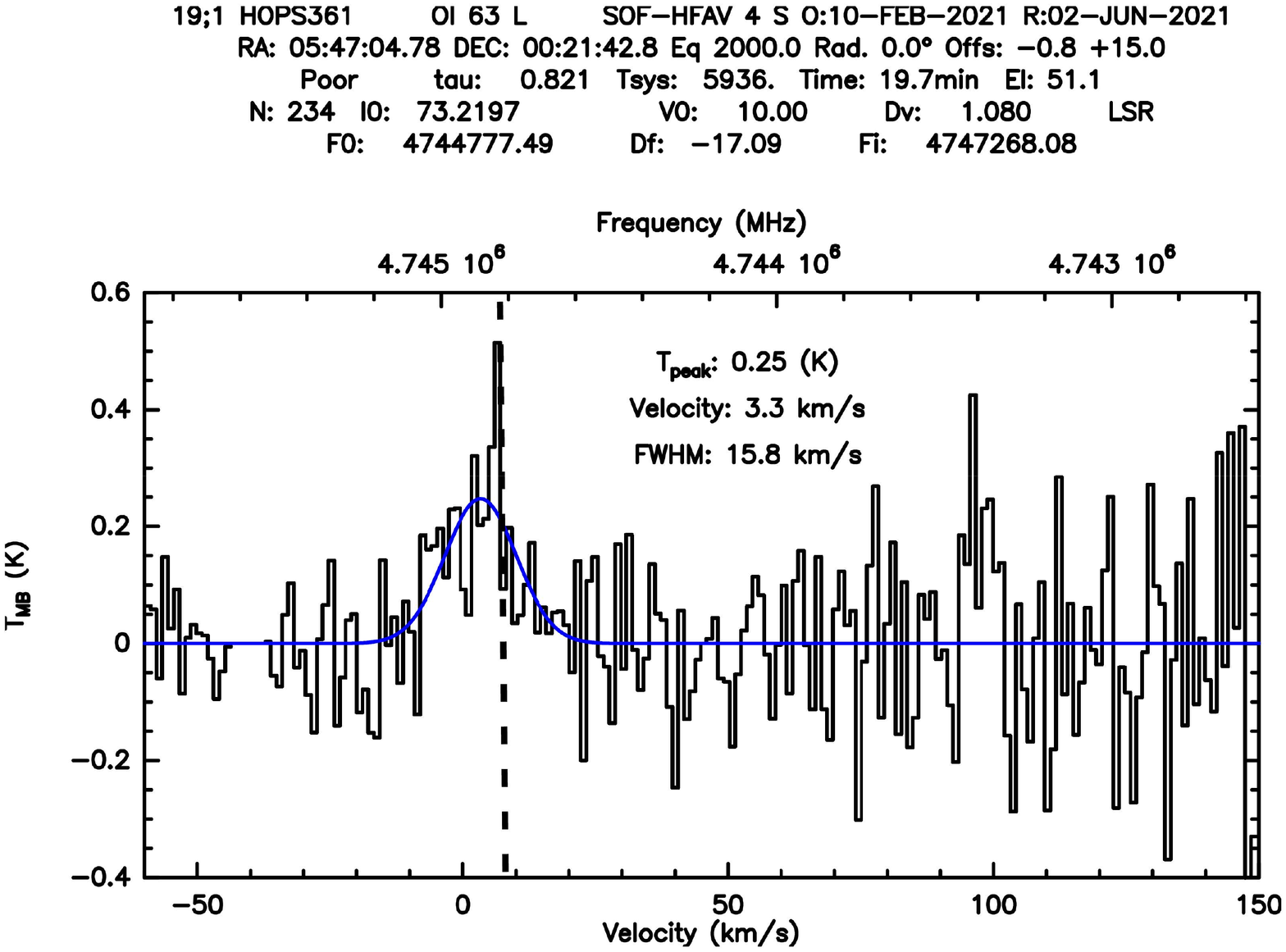}
        \caption{The SOFIA 4GREAT spectrum in black, centered on the [O I] 63 \mic \, line, and positioned near the base of the HOPS 361-C jet. For the footprint on the sky, see Figure \ref{fig:compare_epochs}. The blue curve marks a Gaussian fit to the line centered at around 3.3 km/s with parameters listed in the figure. The dashed line shows HOPS 361-C's systemic velocity of 9.5 km/s for reference \citep{Cheng_2022}. 
        }
        \label{fig:sofia_spectrum}
\end{figure*}

There is only one pointing that overlaps the base of the HOPS 361-C jet (see the circular region centered on the RA and Dec of 05h47m04.723s, 00d21m57.84s in Figure \ref{fig:compare_epochs}). The footprint is shown with a diffraction limited beam size with a radius of 3.15 arcsec. After removing the second order baseline, the resulting [O I] 63 \mic\ line spectrum is shown in Figure \ref{fig:sofia_spectrum} with a Gaussian profile that is fit by least squares minimization. The fit gives a mean velocity of $3.3 \pm 0.5$ km/s and a full width at half the maximum value (FWHM) of $15.8 \pm 2$ km/s. Comparing the mean velocity with the HOPS 361-C protostar's systemic velocity of 9.5 km/s \citet{Cheng_2022} yields an average radial velocity at the base of the jet of approximately -6.2 km/s and a velocity spread with FWHM of about 16 km/s.

    \label{great_obs}


\section{Identifying Knots and Proper Motion Methods} 
\label{analysis_promot}
    We used the HST images to identify bright knots of ionized gas in the HOPS 361-C jet and measure their proper motions tangential to the line of sight between the first and second epoch. The images from the two epochs were taken with different filters, so we constructed a synthetic broadband F160W image for the second epoch. We use the second epoch [Fe II] F164N image, representative of line emission, and the F167N image, corresponding to the continuum, to construct the synthetic F160W image. We set intensities below 0 and above ${10}^{-14}$ to 0 to clip data outside typical knot intensities. We then scale the first epoch image by 9.16 $\times {10}^{15}$, the second epoch image by 8.53 $\times {10}^{14}$, and added 0.168 to the difference. Other linear combinations give similar morphology, but this combination allows the range of intensities for epoch 1 and 2 to be negative and positive respectively.

Our synthetic epoch 2 F160W image is shown in the top right of Figure \ref{fig:compare_epochs}. The Figure \ref{fig:compare_epochs} bottom panel shows the difference between the first epoch F160W image and the synthetic image from the second epoch. We use this difference image to identify the center of bright knots and measure their proper motions between epochs.

The centers for fast-moving knots are visually identified from regions with strong gradients in the difference image (i.e., white-green or blue-green in Figure \ref{fig:compare_epochs}). Ellipses centered on each knot are adjusted by hand in size and orientation by blinking between the epochs (the upper panels of Figure \ref{fig:compare_epochs}) using {\tt{SAOImage DS9}} software. In Table \ref{tab:knot_measures}, we show our elliptical regions for each knot, including their coordinates in each epoch as well as their dimensions. For zoomed-in versions of these knots, see Figure \ref{fig:app_zoomin} in the Appendix (Section \ref{sect: Appendix}).



\begin{deluxetable*}{cccccccc}
    \tablewidth{0pt}
    \tablecaption{Knot Measurements with Regions}
    \tablehead{
    \colhead{Knot Identifier} & \colhead{Epoch 1 RA, Dec}  & \colhead{Epoch 2 RA, Dec} & \colhead{Semi-Major Axis} & \colhead{Semi-Minor Axis}  &  \colhead{\Av} \\
    \colhead{...} & \colhead{(dd:mm:ss, hh:mm:ss)}  &    \colhead{(dd:mm:ss, hh:mm:ss)} & \colhead{(arcsec)}  &    \colhead{(arcsec)}  &  \colhead{(mag)}
    }
    \startdata
        1    &   5h47m04.7278s, +0d21m54.424s & 5h47m04.7749s, +0d21m56.281s &                     1.98 &                     0.75 & 30.3 $\pm$ 1.63 \\
        2    &   5h47m04.9808s, +0d22m05.917s & 5h47m04.9869s, +0d22m07.256s &                     0.32 &                     0.22 & 24.0 $\pm$ 1.48 \\
        3    &   5h47m04.9050s, +0d22m15.608s & 5h47m04.9488s, +0d22m17.079s &                     2.48 &                     0.97 &  9.6 $\pm$ 3.85 \\
        4    &   5h47m05.2484s, +0d22m27.308s & 5h47m05.2577s, +0d22m28.261s &                     0.81 &                     0.45 & 16.0 $\pm$ 0.73 \\
        5    &   5h47m05.3264s, +0d22m28.448s & 5h47m05.3562s, +0d22m29.181s &                     0.83 &                     0.49 & 17.1 $\pm$ 0.67 \\
        6    &   5h47m05.4630s, +0d22m30.251s & 5h47m05.4940s, +0d22m30.948s &                     0.79 &                     0.71 & 20.3 $\pm$ 0.97 \\
        7    &   5h47m05.8456s, +0d22m37.350s & 5h47m05.8837s, +0d22m37.836s &                     1.12 &                     0.65 &  30.7 $\pm$ 1.8 \\
        8    &   5h47m05.9158s, +0d22m39.970s & 5h47m05.9315s, +0d22m40.507s &                     0.62 &                     0.56 & 31.4 $\pm$ 1.13 \\
        9    &   5h47m06.6270s, +0d22m44.653s & 5h47m06.6131s, +0d22m45.252s &                     0.93 &                     0.92 & 20.8 $\pm$ 2.04 \\
        10    &   5h47m08.2013s, +0d22m53.882s & 5h47m08.2241s, +0d22m53.988s &                     3.70 &                     1.26 &  7.7 $\pm$ 8.78 \\
        11    &   5h47m08.4797s, +0d22m54.173s & 5h47m08.4879s, +0d22m54.500s &                     1.69 &                     1.06 &  1.9 $\pm$ 8.75 \\
    \enddata
    \tablecomments{Knot measurements include central knot coordinates and the ellipse region parameters for each knot. The knot identifiers are number labels for knots, where knot 1 is closest to HOPS 361-C, and knot 11 is the furthest knot. We also show the \textbf{mean} extinction, \Av, for each knot with their respective median uncertainties in Figure \ref{fig:speed_trends}.
     \label{tab:knot_measures}}
      \vspace{-0.3in}
\end{deluxetable*}

\begin{figure*}
    \includegraphics[width=0.95\textwidth]{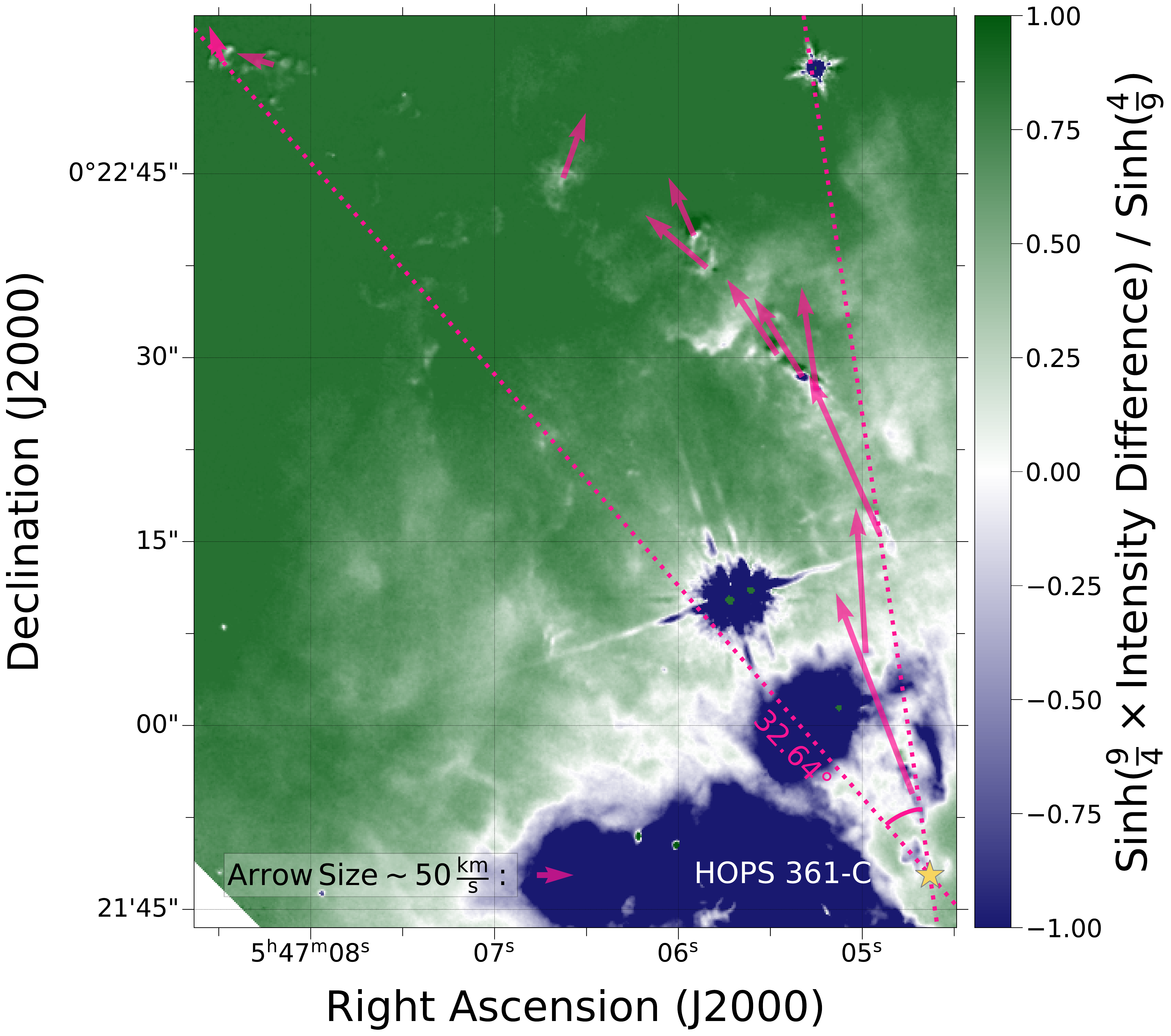}
    \caption{A zoom-in of the difference image from Figure \ref{fig:compare_epochs} with pink arrows made from our proper motion measurements (listed in Table \ref{tab:propmot}). The yellow star marker in the bottom right corner is HOPS 361-C, which we assume is the protostellar source for the jet. The labeled angle and dotted lines mark the knots from HOPS 361-C that form an extreme angle. }
    \label{fig:arrow_image}
\end{figure*}

We measure proper motions by computing how many pixels each knot’s center shifts between epochs. Shifts in units of pixels listed in Table \ref{tab:knot_measures} are converted to angular sizes using HST's pixel size (0.12825 arcsec) and to physical lengths using the adopted distance of 430.4 pc from \citet{Tobin_2020}. To find speeds, we divide by the time between epochs (11 yrs). Table \ref{tab:propmot} lists the shifts in arcsecs and the resulting proper motions. Figure \ref{fig:arrow_image} illustrates the direction of motion with arrows at the location of each knot.   

\begin{deluxetable*}{cccccccc}
    \tablewidth{0pt}
    \tablecaption{Knot Motion and Speeds}
    \tablehead{
    \colhead{Knot Identifier} &
    \colhead{RA Shift}  & \colhead{Dec Shift}  & \colhead{Proper Motion}  & \colhead{Tangential Speed \textbf{($v_P$)}}  & \colhead{Position Angle}  & \colhead{Shock Speed \textbf{($v_S$)}} &  \colhead{Total Outflow Speed \textbf{($v_{outflow}$)}}   \\
    \colhead{...} & \colhead{(arcsec)}  &    \colhead{(arcsec)}  &    \colhead{(arcsec/yr)}  &    \colhead{(km/s)}   & \colhead{(degrees)}  &    \colhead{(km/s)}  &    \colhead{(km/s)}
                }
    \startdata
        1    &   0.689 &      1.857 &          0.180 &             290.7 &            20.8 &         51.6 &           342 \\
        2    &   0.078 &     1.339 &          0.122 &             196.6 &             3.9 &         50.2 &           247 \\
        3    &   0.643 &     1.471 &          0.146 &             235.6 &            24.0 &         46.9 &           282 \\
        4    &   0.130 &     0.953 &          0.087 &             141.2 &             8.3 &         50.9 &           192 \\
        5    &   0.440 &     0.733 &          0.078 &             125.5 &            31.4 &         49.1 &           174 \\
        6    &   0.458 &     0.697 &          0.076 &             122.4 &            33.7 &         52.3 &           174 \\
        7    &   0.567 &     0.486 &          0.068 &              109.6 &            49.6 &         52.6 &           162 \\
        8    &   0.230 &     0.537 &          0.053 &             85.8 &            23.7 &         47.8 &           133 \\
        9    &   -0.214 &     0.599 &          0.058 &             93.4 &           -19.2 &         49.3 &           142 \\
        10    &  0.341 &    0.106 &          0.032 &              52.4 &            72.8 &         49.8 &            102 \\
        11    &  0.120 &    0.327 &          0.032 &              51.1 &            20.6 &         48.5 &            99.6 \\
    \enddata
    \tablecomments{Motion-based values derived for each knot, including how much knot centers move between epochs (see Figure \ref{fig:app_zoomin}), proper motions, and tangential speeds. The knot identifiers are shown as in Table \ref{tab:knot_measures}, where 1 is the knot closest to HOPS 361-C, and 11 is the furthest. Position angles for the tangential velocity vectors are measured between HOPS 361-C and each knot, relative to celestial North. Shock speeds added with the tangential speeds result in the total flow speed through each knot.
     \label{tab:propmot}}
     \vspace{-0.2in}
\end{deluxetable*}

Proper motions for Herbig-Haro (HH) objects can be determined using the centroid for a box surrounding a knot of interest \citep{Bally_2002, Reiter_2017, Erkal_2021}, shifting one epoch relative to the next and minimizing the square of the difference summed over a box \citep{Hartigan_2001}, by fitting more complex shapes (e.g. a symmetric parabola or Gaussian) to a knot or the jet \citep{Eisloeffel_1992}, or by cross-correlation of both images \citep{Reipurth_1992, Raga_2012, Raga_2017}. We attempted differently-shaped regions, centroids, and phase-based cross correlations but found these methods sensitive to noise, thresholding, and background subtraction. Automated methods are hampered by the different filters between the two epochs and because non-rigid knots broke up or changed shape between epochs. We find that our results are consistent with other methods but produce less noisy knot centers.

Figure \ref{fig:arrow_image} shows that the proper motions and arc of knots overall appear consistent with outflows originating from HOPS 361-C. Our result is concordant with prior studies that found proper motion directions tend to align with a vector to their source \citep[e.g.][]{Anglada_2007}. But it is unclear if these knots all originate from the HOPS 361-C jet. To check, we determine extinction to infer each knot's relative location within the cloud, and we confirm whether these knots have shock properties akin to other protostellar jets.

\section{Extinction and Shock Modeling} 
\label{analysis_shocks}

   \subsection{Observed Extinctions} 
    We need spatially-varying extinction measurements in order to interpret line ratios in terms of shock models and to estimate how deeply a protostar and its jet are embedded within the molecular cloud. We estimate extinction with the second epoch [Fe II] 1.26 \mic~and [Fe II] 1.64 \mic~spectral line images. These lines share a common upper state, so their intrinsic or unextinguished ratio can be computed from the wavelengths and Einstein A-coefficients for the transitions. The primary uncertainty is that the Einstein A-coefficients are difficult to determine theoretically.

\begin{figure*}
    \centering
    \includegraphics[width=\textwidth, trim = 0in 0in 0in 0in,clip]{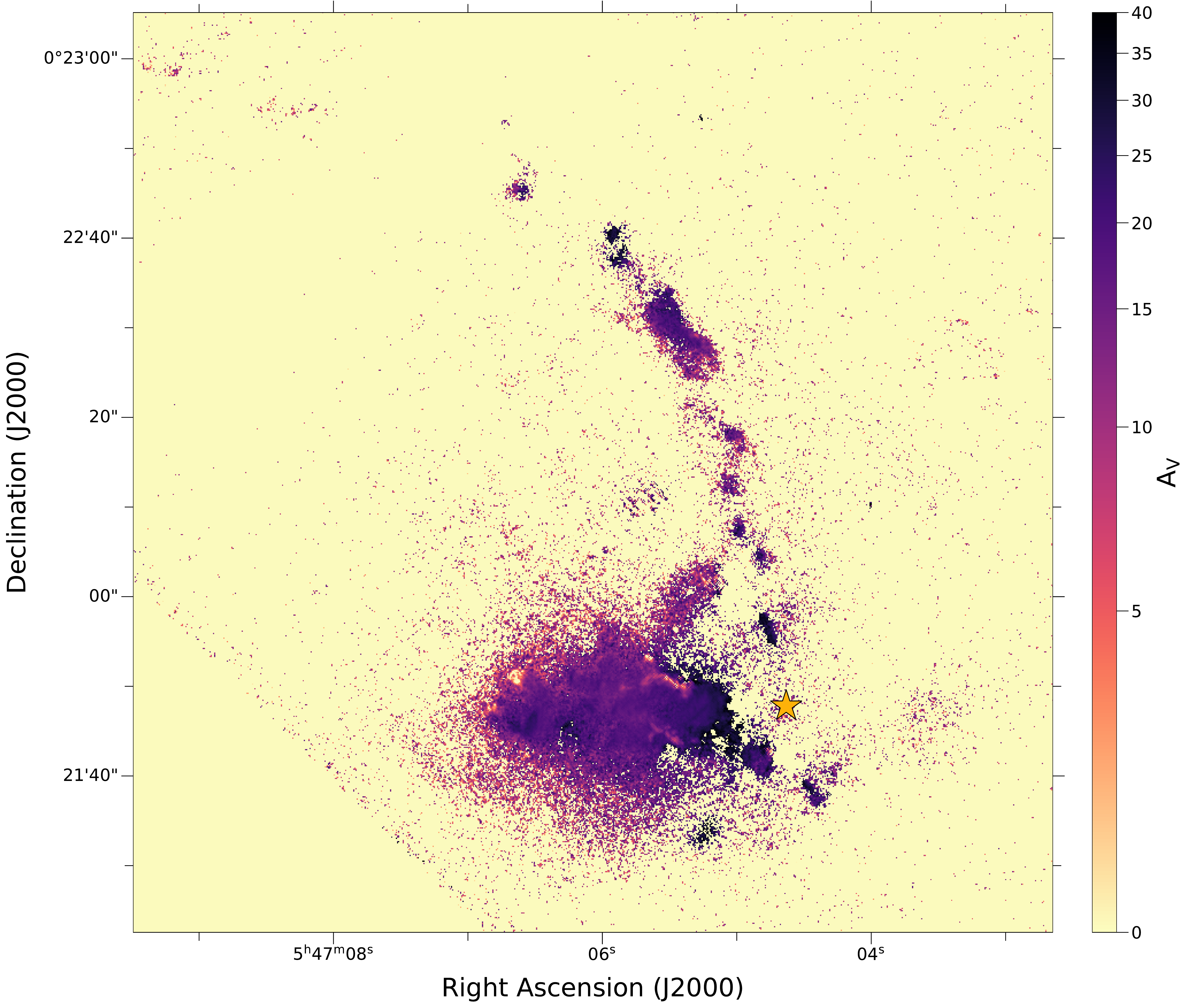}
    \caption{\Av \ map \ of HOPS 361 with an arcsinh stretch. Values are computed using the [Fe II] 1.26 \mic/[Fe II] 1.64 \mic~spectral line intensity ratio, with each line image having an SNR $>$2. HOPS 361-C is shown as a yellow, star-shaped marker. }\label{fig:av}
\end{figure*}

We mitigate uncertainties in A-coefficients by adopting an observed zero-reddening value of 2.6 for the [Fe II] line intensity ratio based on observations of bright HH objects in NGC 1333 \citep{Rubinstein_2021}. Next, we use the $R_V=5.5$ model by \citet{Weingartner_2001} to find an optical depth from the ratio of the observed and theoretical [Fe II] line intensity ratios. The resulting map of \Av \ for the HOPS 361 region is shown in Figure \ref{fig:av}. A signal-to-noise (SNR) ratio of 2 is used to clip any signal in either HST line image below twice the noise of $1.55 \times {10}^{-18}\ {\rm erg}\ {\rm cm}^{-2}\ {\rm s}^{-1}$ per pixel from Section \ref{hst_obs}.

We use the regions we define in Section \ref{analysis_promot} to estimate the mean extinction to each knot, both of which are listed in Table \ref{tab:knot_measures}. For uncertainties, we use the 'error' image or uncertainty plane directly outputted by {\tt drizzlepac} for each image. The uncertainty for each pixel consists of exposure times taking account of Poisson noise (e.g. shot noise), read noise, and other sources of uncertainty. We propagate through the steps of calculating line ratios and extinction. The uncertainties show extinction is well-estimated in the vicinity of HOPS 361-C and the arced jet where we detect proper motions, until the end of the jet where extinction drops off.

    \label{ext} 
    
    \subsection{Shock Speed and Pre-Shock Density} 

Emission-line knots along the HOPS 361-C outflow are HH objects: ionic-line emission follows the passage of a dissociative and ionizing shock produced by supersonic variations in outflow speed or the outflow supersonically interacting with the ambient medium. At each knot, an outflow, initially launched at speed $v_j$ and pre-shock hydrogen number density $n_0$, enters a shock at relative speed or shock speed $v_S$. It decelerates and compresses in a (Rankine-Hugoniot) jump to speed $v_S /4$ and density $4n_0$, then to a lower speed and higher density determined by momentum conservation and radiative cooling. 

Post-shock velocity, the vector sum of proper motion $v_{P}$ and radial velocity $v_{R}$, differs from outflow velocity $v_{outflow}$ by $v_S$. Post-shock ionic-line emission intensities are sensitive to $v_S$ and $n_0$, so $v_{P}$ and $v_{R}$ measurements enable us to determine outflow speed $v_{outflow}~\approx~v_S~+~v_{P}$, mass density $\rho_0 = \mu n_0$ ($\mu$ as the mean molecular weight), mass flux $\rho_0 v = \mu n_0 v$, momentum flux $\rho_0 v^2$, and kinetic energy flux $\rho_0 v^3$ lost by the outflow at every knot.   

We determine $v_S$ and $n_0$ from observed line emission with the 1-D shock and photoionization code MAPPINGS V \citep{Sutherland_2018}. For [Fe~II], we substitute the network of [Fe~II] radiative and collisional rate coefficients calculated by \cite{Tayal_2018}, but otherwise use the default atomic-physical data from the CHIANTI database \citep{CHIANTI}. Typical protostellar shock speeds are small enough that little sputtering of dust is expected, so we assume the same gas-phase abundances as \cite{Watson_2016}. We generate a grid of model line-emission spectra with $v_S = 20-62~\rm{}km~s^{-1}$ and $n_0~=~0.1-320,000~\rm{cm^{-3}}$. At each grid point we assume a magnetic field oriented perpendicular to the outflow and frozen into the medium at typical interstellar strength, $B~=~5~\mu G\sqrt{n_0/100~\rm{cm}^{-3}}$. Each calculation terminates at gas temperature $T~=~100~\rm{K}$. While MAPPINGS is 1-D, HH objects are well-resolved structures observed close to edge-on, so fits to observed intensity can still estimate $v_S$ and $n_0$.
 
\begin{figure}
    \centering   
      \includegraphics[width=\columnwidth, trim = 0in 0.5in 0in 0in,clip]{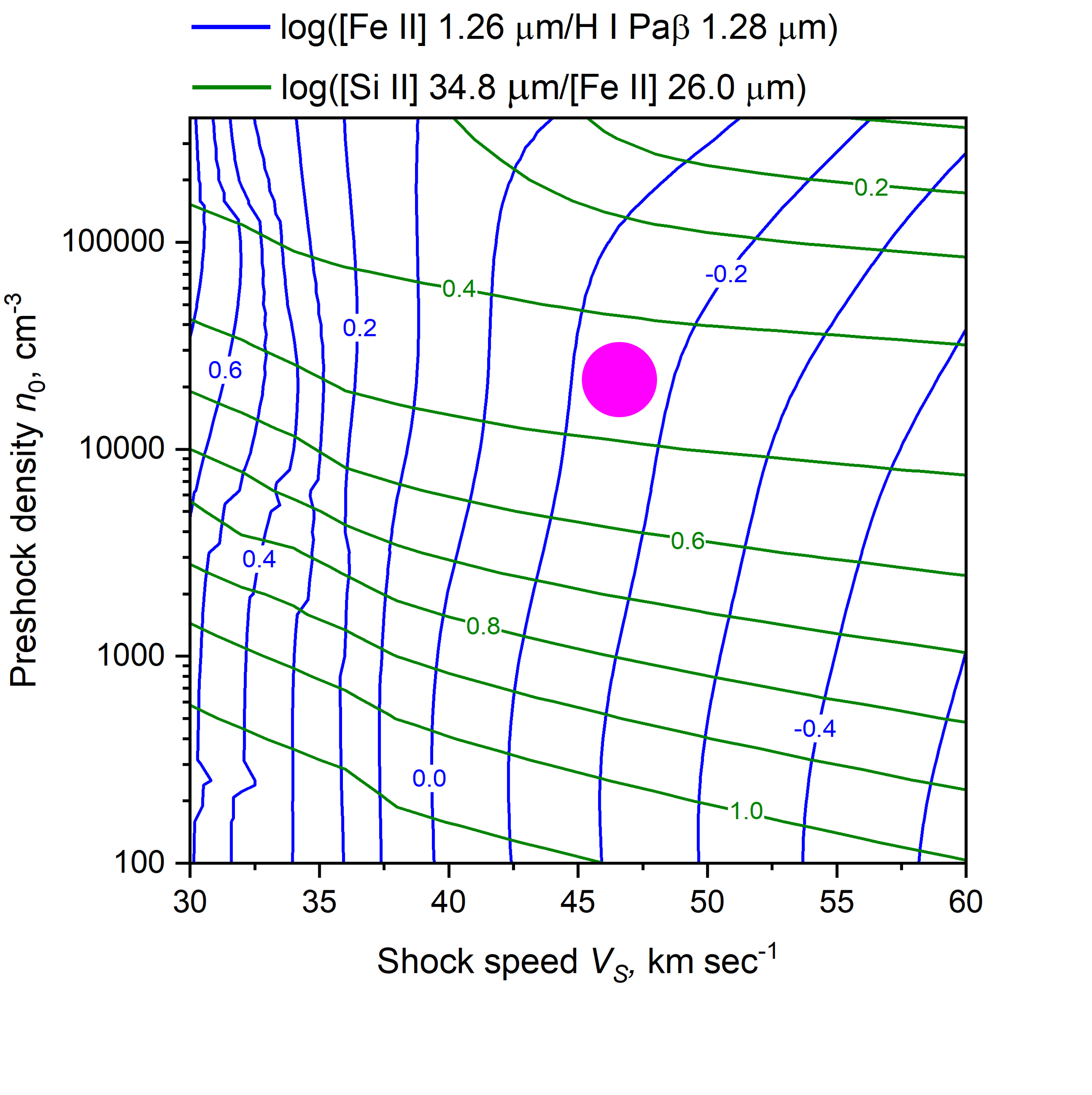}
    \caption{
    Line ratio nomogram from the MAPPINGS V model grid: log([Fe~II] 1.26~\mic /HI~Pa$\beta$ 1.28~\mic ) in blue,  log([Fe~II] 26.0~\mic /[Si~II] 34.88~\mic ) in green, plotted against shock speed $v_S$ and preshock density $n_0$. The magenta circle indicates the locus of line ratio values with well-defined HH objects in both the HST and Spitzer spectroscopic images, It implies shock speed $v_S = 45.7~\rm{km~sec^{-1}}$ and preshock density $n_0 = 3.2\times 10^4 ~\rm{cm^{-3}}$. 
    }\label{fig:nomogram}
\end{figure}

Our model-grid results, derived parameters, and sensitivity to observations are in Figure \ref{fig:nomogram}. Intensity ratios are from our HST images. Blue, nearly vertical contour lines can be a ratio of HI~Pa$\beta~1.28~\mu\rm{m}$ to either NIR [Fe~II] line. This line ratio is sensitive to $v_S$ and relatively insensitive to $n_0$. Green, nearly horizontal contours show the ratio of mid-infrared [Si~II] 34.8 \mic~and [Fe~II] 26.0 \mic~emission lines. This ratio is sensitive to $n_0$ and insensitive to $v_S$. The northern half of the jet, imaged with the Spitzer Infrared Spectrograph at 15 arcsec spatial resolution, includes the spectral lines [Fe~II]~26.0~\mic \ and [Si~II]~34.8~\mic \ \citep{Melnick_2008}. 

Three bright regions in the Spitzer images (labelled P1, P2, and P3 in Figure 6 of \citealt{Melnick_2008}) are chosen to measure the [Si II] 34.8 \mic\ to [Fe~II] 26.0 \mic\ line ratio of $\sim$0.45. Peaks in the HST images in the same regions are used to measure the ratio of [Fe~II] 1.26 \mic\ to Pa$\beta$, about -0.15. These two line ratios are placed on the MAPPINGS shock model in Figure \ref{fig:nomogram} with a pink circle whose size approximates the scatter from differences of ratios at the different peaks. From the location of the pink dot in Figure \ref{fig:nomogram}, we estimate $n_0~\approx~3.2 \times 10^4~\rm{cm}^{-3}$ and $v_S~=~40-50~\rm{km~sec}^{-1}$. 

\begin{figure*}
    \centering   
    \includegraphics[width=\textwidth, trim = 0in 0in 0in 0in,clip]{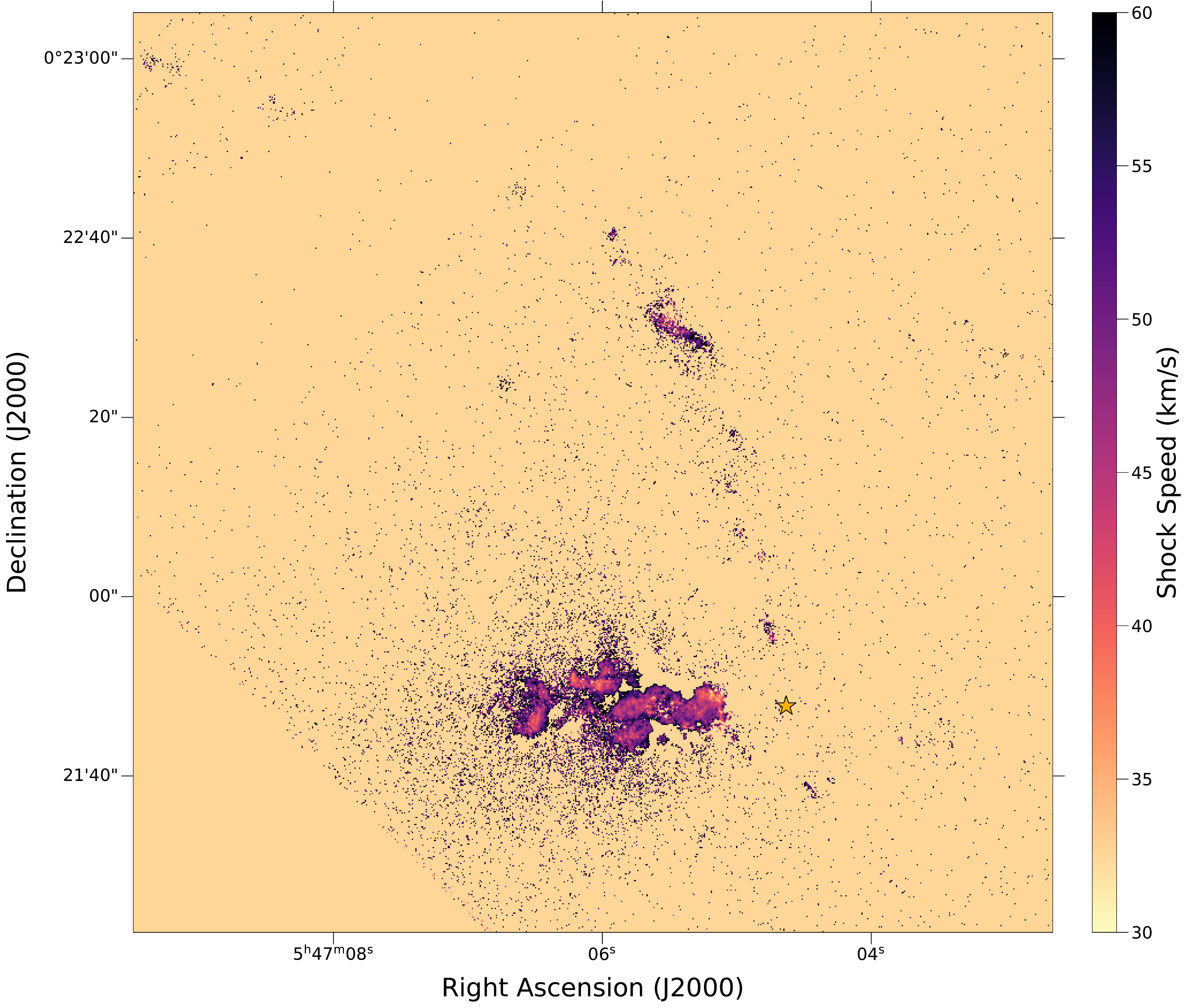}
    \caption{
    The shock speed map computed from the [Fe II] 1.64 \mic/H I Pa $\beta$ 1.28 \mic~intensity ratio with each line image having an SNR of 1 and using Equation 1. HOPS 361-C is shown as a yellow, star-shaped marker.
    }\label{fig:shocks}
\end{figure*}

HST images may show morphology not evident in relatively lower resolution Spitzer images. We use the [Fe II] 1.26 \mic\ to Pa$\beta$ line ratio, computed from the HST images alone, to create an image that shows the shock speed derived from the MAPPINGS model grid. Taking advantage of the density insensitivity for the NIR [Fe~II]/HI~Pa$\beta$ ratio, we fit a third-order polynomial to our grid points to estimate shock speed which is accurate to a few $\rm{km~sec}^{-1}$: 
\begin{align} \label{eqn:VS} 
 v_S(R) &= -12.617 {\log_{10}{R}}^3 + 30.684 {\log_{10}{R}}^2 \\
 &\quad - 42.044 \log_{10}{R}+33.222, \notag
 \end{align}
where $R$ is the [Fe II] 1.26~\mic/Pa$\beta$~1.28~\mic\ intensity ratio. We use this relation to generate the shock-speed image in the bottom panel of Figure \ref{fig:shocks} with each line imge at an SNR above 1. We neglect correcting the [Fe II] 1.26 \mic \ and Pa$\beta$ lines for extinction. The two lines are close in wavelength so the effect of extinction is negligible. 

Unlike proper motions, shock speeds do not appear to depend strongly on distance along the outflow from HOPS 361-C. The shock speed map in Figure \ref{fig:shocks} shows the path of knots according to our proper motions may extend further than about 0.2 pc from HOPS 361-C, following bright trails of molecular hydrogen emission \citep{Walther_2019, Eisloeffel_2000} and extended line emission in Figure \ref{fig:mosaics_epoch2}. Making conclusions about into neighboring frames is difficult without detecting proper motions and bright knots to the southwest of the HOPS 361 clump. There appears to be a strip of shocked material neighboring HOPS 361-E, but the source is unclear in the first epoch. According to our extinction map in Figure \ref{fig:av}, this part of the HOPS 361 clump has a higher degree of extinction. Ambiguity remains as the trail may consist of a single arc or form a longer wave as the result of previous periodic ejections in a precessing jet. 

    \label{jetmod} 

\section{Discussion} 
\label{discussion}  
    \subsection{HOPS 361-C Jet Speed} 
Using shock speeds (estimated via equation \ref{eqn:VS}) and proper motions, we estimate the total outflow speed through each knot. We then examine how outflow speed depends on distance from the HOPS 361-C source. A decrease in flow velocity as a function of distance from the source may be consistent with a jet slowed or decelerated by the host molecular cloud. Such a jet may inject its energy and momentum into its host molecular cloud and locally affect stellar feedback. 

To estimate total flow speed, following \citet{Coffey_2004} and \citet{Bally_2002}, we sum the 3D velocities of post-shock gas, as measured by the proper motions and radial velocity added in quadrature, with the velocity jump through the shock from Section \ref{jetmod}. We ignore radial velocity because the majority of the speed is due to the 100--350 km/s tangential motions as opposed to the radial velocity of $<$5 km/s, so these velocities indicate that the jet is oriented close to the plane of the sky \citep[for examples exploring protostellar jets, their resultant shocks, and inclination effects, see][]{Hartigan_2000, Graham_2003, Jhan_2022}. According to the 1-D shock model described in Section \ref{jetmod}, half of the velocity spread of shocked material measured from [O I] in Figure \ref{fig:sofia_spectrum} should be between the speed of turbulent gas measured from C$^{18}$O \citep[1 km/s, see][]{Iwata_1988, Stanke_2022} and roughly a quarter of the shock speed (10 km/s). With a velocity spread of 6.2 km/s, our assumed radial velocity and low inclination are consistent with the model.

\begin{figure*}
    \centering
    \includegraphics[width=\textwidth,height=0.94\textheight,keepaspectratio]{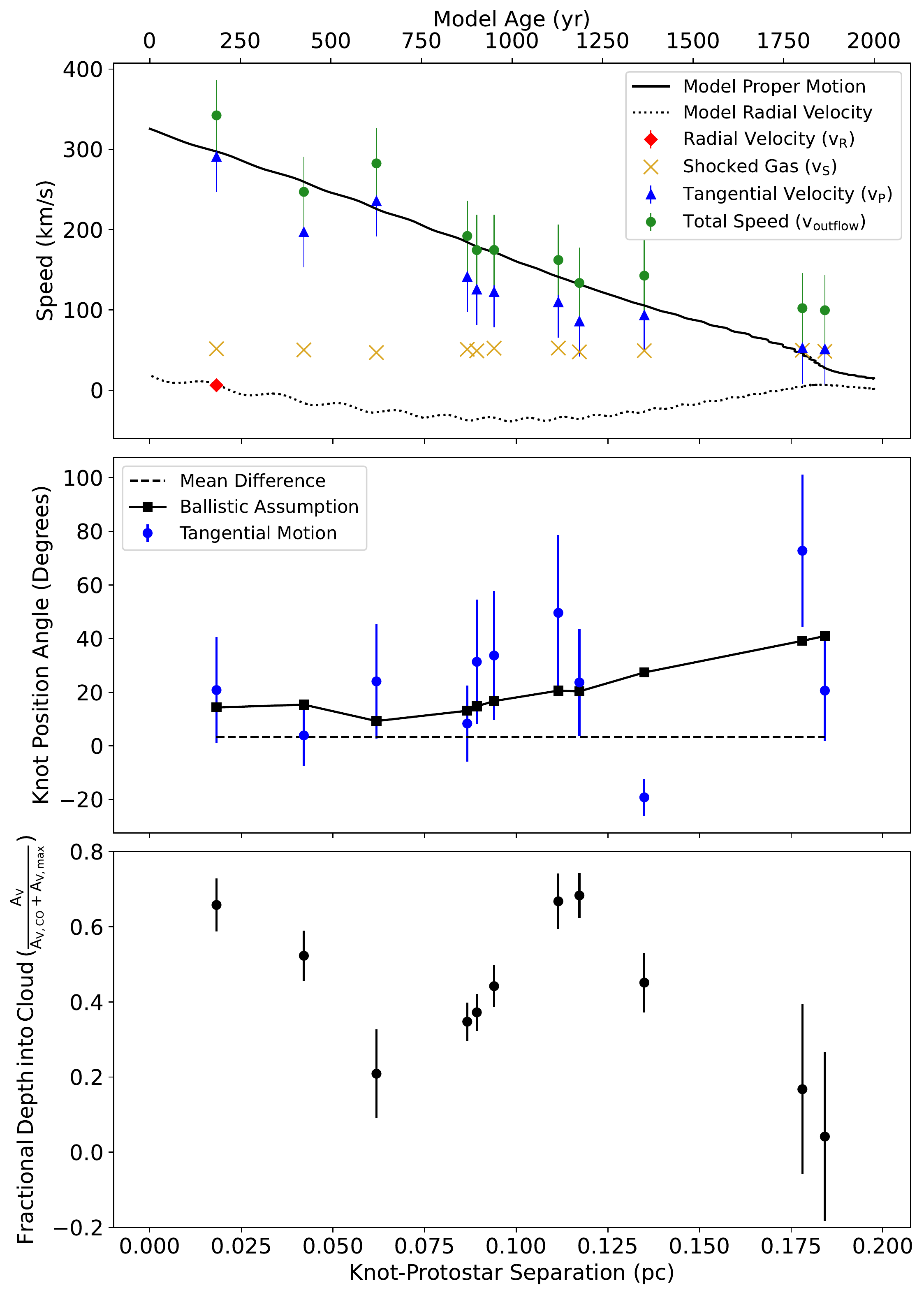}
    \caption{\underline{\textit{Top}:} The tangential speed derived from proper motions (blue triangles), shock speed from the [Fe II]/Pa$\beta$ line ratio (yellow x's), and bulk speed of the jet and outflow through each knot from adding the two (green dots). A precession model is plotted for the tangential speed (solid curve) and radial velocity (dotted curve). The red diamond shows the observed radial velocity for knot 1 estimated from the [O I] 63 \mic\ spectrum. The modeled time since ejection is shown on top of the plot. \\
    %
    \underline{\textit{Middle}:} Proper motion position angles relative to celestial North (blue dots), and expected position angles for knots ballistically ejected by HOPS 361-C (black squares). 
    The dashed line is the 4$^\circ$ mean difference between observed and expected angles. \\
    \underline{\textit{Bottom}}: We estimate each knot's depth within the cloud from the ratio of extinction to the knot and the cloud's total extinction, based on  $\rm C^{18}$O observations. Here, the $x$-axis shows the direct distance on the sky to the likely protostellar source, HOPS 361-C. 
    }\label{fig:speed_trends}
\end{figure*}

In Figure \ref{fig:speed_trends} we show quantities for each knot listed in Table \ref{tab:propmot} as a function of distance from the likely protostellar source, HOPS 361-C. In the top panel, blue triangles show tangential velocity directly found from the proper motions. Yellow x's show the velocity of the shocked gas, estimated from the [Fe II]/Pa$\beta$ line ratio. The green dots show the sum of these two quantities, which is an estimate for each knot's total outflow velocity. The red diamond shows the radial velocity for the jet from our SOFIA spectrum in Figure \ref{fig:sofia_spectrum}. 

The top panel of Figure \ref{fig:speed_trends} shows that the outflow velocity is decelerating as a function of distance from the source. Since proper motions dominate the speed, their uncertainty values are plotted by applying deviations of 2 pixel, from the angular resolution or PSF's FWHM of epoch 1, to the offsets listed in Table \ref{tab:propmot}, which corresponds to a speed of about 44 km/s. The velocity drops from about 350 km/s to about 100 km/s over a distance of 0.2 pc, which may be because of collisions in a dense environment \citep[e.g.][]{Velazquez_2013}. 

The amount of deceleration is relatively larger than that in more linear jets suspected to also impact into dense gas. For example, HH 1 is noted to have almost no deceleration ($<$50 km/s) over a distance of approximately 0.15 pc despite interactions within a dense clump \citep{CastellanosRamirez_2018, Raga_2017}. HH 2 is similarly noted to not have much or any deceleration over a distance of 0.02 pc after finding low dispersion in over 70 years of data \citep{Raga_2016b}.

Jets have knot velocities typically aligned with the direction of their source \citep[e.g.][]{Schwartz_1999, Heathcote_1996, Raga_1993}. In the middle panel of Figure \ref{fig:speed_trends} the blue points show the direction of motion for each knot relative to celestial North, based on the proper motions. Black squares show the position angle of the vector between each knot and HOPS 361-C. The uncertainties plotted for the proper motions apply the same worst case scenario using 2-pixel deviations in our measured offsets. The observed velocity vectors have position angles that are about 4$^\circ$ higher than what is predicted if the knots were shot out ballistically (i.e. a straight line on the sky) from HOPS 361-C. 
That said, our uncertainties cannot rule out ballistic motion on average for the majority of knots except the outlier at approximately 0.13 pc.

In case the discrepancy between the proper motion’s direction and direction to source is real, potential explanations are that the knots are launched by a different protostellar source than HOPS 361-C, a projection effect (e.g. more extreme radial velocities, another protostar launching the knot), or non-ballistic motion.  
We searched for an alternative source but did not find a candidate among the known HOPS 361 protostars (A--E) and the nearby protostars HOPS 335 and HOPS 366.
Deviation from ballistic motion for the outlier at 0.13 pc could be caused by deflection due to dense clumps of gas in the ambient molecular cloud material, which may change the direction of a knot \citep{Raga_1995}.

    \label{outflowspeed}

   \subsection{A Decelerating Jet Precession Model}
To match the positions and proper motions of our knots of ionized gas, we construct a model similar to those used to model wiggles observed in jets. For background and diagrams of precessing jets, please see \citet{Masciadri_2002, Anglada_2007}.

We adapt the analytical model from \citet{Raga_2009} and invoking constant jet ejection velocity \citep{Raga_1993} but with an additional damping term (i.e. time-dependent exponential decay) to account for the observed decreasing speeds in Figure \ref{fig:speed_trends}. Constant density jets with a decaying, time-dependent velocity profile are not without precedent in jet models \citep[e.g.][]{Kofman_1992, Velazquez_2013}, modeling observed radial velocities and proper motions for HH objects (e.g. HH 34 by \citealt{Cabrit_2000}, HH 80/81/80N by \citealt{Masque_2015}, HH 223 by \citealt{Lopez_2015}), and modeling knots of CO gas for a moving protostellar source \citep[e.g. PV Cep by ][]{Goodman_2004}.

We consider a binary comprised of
two stars with masses $m_1$ and $m_2$ in a circular orbit of radius $a_B$. 
The jet is assumed emit from $m_1$ and the binary orbit lies in the $x,y$ plane.
Due to the motion of $m_1$ about the center of mass of the binary system, 
the initial position from the jet source as a function of ejection time $t_e$ is
\begin{align}
x(t_e) &= r_0 \cos (\omega_B t_e)\nonumber \\
y(t_e) &= r_0 \sin (\omega_B t_e) 
\end{align}
where the binary system's mean angular motion is 
$$\omega_B = \sqrt{\frac{G(m_1 + m_2)}{a_B^3}} .$$ 
Here, the distance from the binary's center of mass position to the jet source, $m_1$, is
\begin{equation}
r_0 = \frac{m_2}{m_1+m_2} a_B .
\end{equation}
The jet precesses if misaligned with the binary orbit's normal \citep{Terquem_1999}.
The precessing jet then has an initial velocity vector originating from $m_1$ 
\begin{equation}
{\bf v}_j(t_e) = v_j ( \sin \beta  \cos (\Omega_p t_e - \phi_0),
\sin \beta  \sin (\Omega_p t_e  - \phi_0),  \cos \beta ) , \label{eqn:vjte}
\end{equation}
where $\phi_0$ is a phase with respect to the binary orbit,  the angle $\beta$ sets the opening angle of the jet with respect to the orbital plane and $v_j$ is the jet velocity.
The jet is assumed to be ballistic so fluid parcels representing knots preserve their velocity at launch. 
Here $\Omega_p$ is the precession rate of the jet. 
Taking into account the orbital motion of the binary, 
the initial velocity vector of jet material emitted at $t_e$ is
\begin{align}
v_x(t_e)  &= -r_0 \omega_B \sin (\omega_B t_e) + v_j  \sin \beta  \cos (\Omega_p t_e - \phi_0) \nonumber \\
v_y(t_e) & = r_0 \omega_B \cos (\omega_B t_e) + v_j  \sin \beta  \sin (\Omega_p t_e - \phi_0) \nonumber \\
v_z(t_e)  &= v_j \cos \beta .
\label{eqn:vte}
\end{align}
We allow the velocity of emitted material to drop as a function of travel time with 
\begin{equation}
{\bf v}(t,t_e) = {\bf v}(t_e) e^{-\alpha (t-t_e)} , \label{eqn:vte2}
\end{equation}
where $\alpha$ describes the decay rate.
Here ${\bf v}(t_e)$ is given in the previous equation. After a knot is emitted, its direction of motion does not change, so the jet propagates ballistically. However, the velocity of emitted material decelerates due to the $\alpha$ parameter. Without damping (in the limit of $\alpha \to 0$), the model reduces to that by \citet{Masciadri_2002} for orbital motion alone. What’s more, if the precession rate is set to zero, or that for precession alone, then binary motion is neglected.

We integrate ${\bf v}(t_e)$ in equation \ref{eqn:vte2} to find the position of
ejected material at a later time $t$
\begin{equation}
{\bf x}(t,t_e) = {\bf x}_0(t_e) + \frac{ {\bf v}(t_e)}{\alpha} (1 -  e^{-\alpha (t-t_e)} ) 
\end{equation}
or specifically
\begin{align}
x(t,t_e) &= v_x(t_e) \frac{1}{\alpha}(1 - e^{-\alpha (t-t_e)}) + r_0 \cos (\omega_B t_e) \nonumber \\
y(t,t_e) &= v_y(t_e)  \frac{1}{\alpha}(1 - e^{-\alpha (t-t_e)}) + r_0 \sin (\omega_B t_e) \nonumber \\
z(t,t_e) &= v_z(t_e)  \frac{1}{\alpha}(1 - e^{-\alpha (t-t_e)}) . \label{eqn:pos}
\end{align}
Here the ejection time is $t_e$ and the present time of observation is $t$. By setting the present time $t=0$, the angle $\phi_0$ sets the jet orientation at the present time.   
We can use equation \ref{eqn:pos} to compute the position of ejected material as a function of ejection time and the velocity of this same material using equation \ref{eqn:vte2}. This gives position and velocity of ejected material in a coordinate system associated with the binary star.   

We rotate along the $x$ axis to correct for the inclination $i_B$ of the binary orbit normal vector.  Here $i_B=0$ corresponds to an edge-on binary orbit plane.  Then we rotate the resulting coordinate system along the line of sight by angle $\xi$ to correct for the position angle of  the binary orbit's normal on the sky.
\begin{equation}
\begin{pmatrix} x_s \\ y_s \\ z_s \end{pmatrix} = 
\begin{pmatrix}
\cos \xi & 0 & -\sin \xi \\
0    & 1 & 0 \\
\sin \xi & 0 & \cos \xi \\
\end{pmatrix}
\begin{pmatrix} 1 & 0 & 0 \\ 0 & \cos i_B & \sin i_B \\   0 & -\sin i_B & \cos i_B \end{pmatrix} 
\begin{pmatrix} x \\ y \\ z \end{pmatrix} 
\label{eqn:rot}
\end{equation}
Coordinates on the sky are $x_s,z_s$,
and $y_s$, 
with $y_s$ increasing away from the viewer along the line of sight.
Here $z_s$ is positive to the north 
and $x_s$ is positive to the west. This is a right-hand coordinate system with origin at the location of the source, HOPS 361-C.  
The rotations given in equation \ref{eqn:rot} are also used to transform the velocity vectors. 
This gives $v_{xs}, v_{zs}$ for
motions in the plane of the sky (tangential motions) and $v_r = v_{ys}$
corresponding to motion along the line of sight (the radial component).
The resulting model has the following free parameters: $v_j, \Omega_p, \alpha, \beta, i_B, \xi, \phi_0, m_1, m_2, a_B$. 

We explore parameter space by hand to see what ranges of parameters match the knot locations, the decrease in outflow velocity as a function of distance from the source protostar and the radial velocity of the first knot. We do not attempt to fit the semi-major axis ($a_B$) or masses of the orbit's constituents ($m_1,m_2$) because these parameters only change the jet trajectory on scales smaller than observed, but we still allow the binary to change the jet's behavior. Lower inclinations relative to the plane of the sky (0--5$^\circ$, see Section \ref{outflowspeed}) narrowed our search through parameter space and better matched positions on the sky, proper motions, and opening angle (Figure \ref{fig:arrow_image}). The first knot’s low radial velocity (Section \ref{great_obs}) and our measured proper motions were better matched with low initial phase angles ($10 < \phi_0 < 20$).
  
\begin{figure*}
    \centering
    \includegraphics[width=\textwidth]{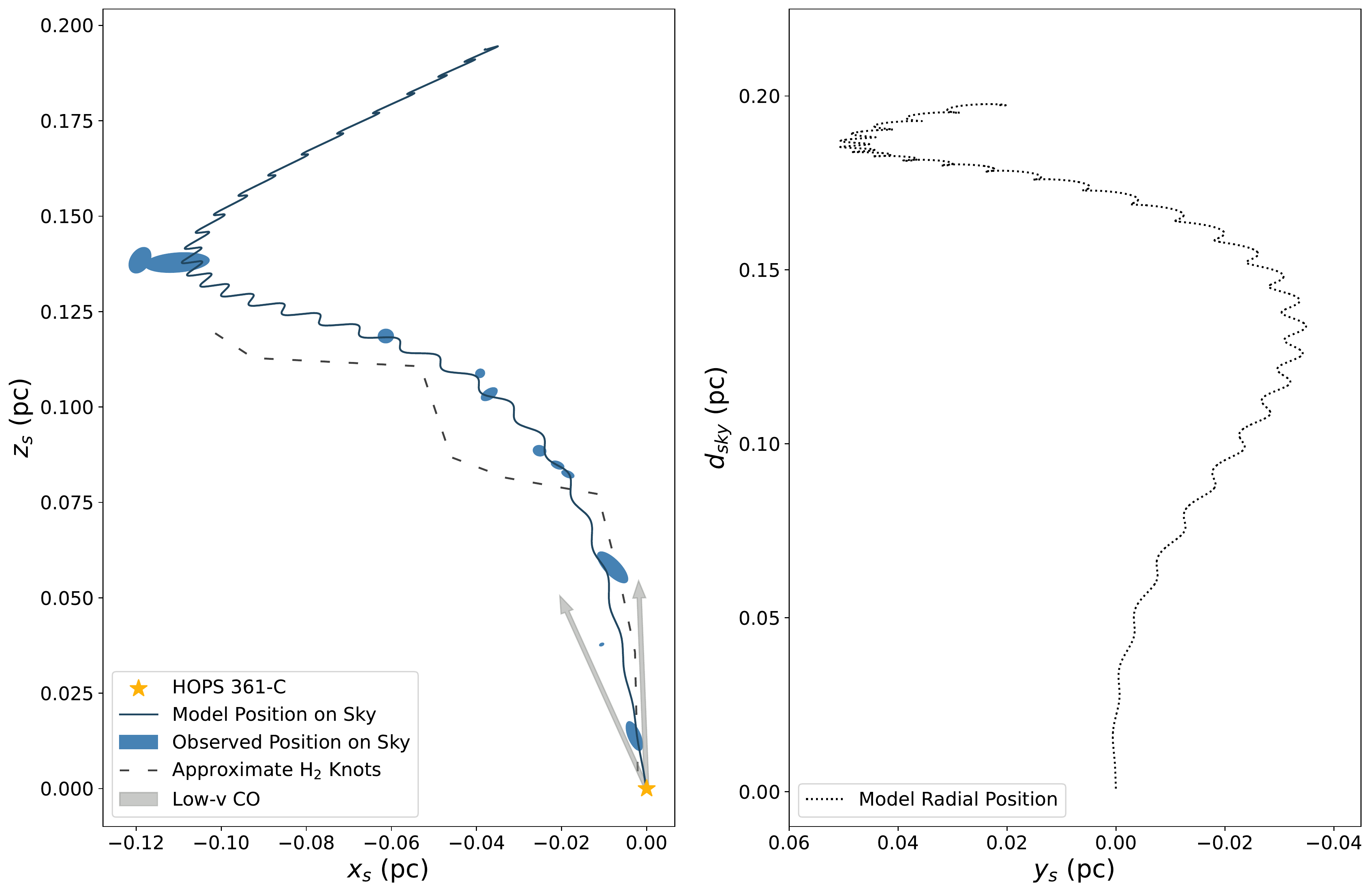}
    \caption{
    \underline{\textit{Left}:} Knot regions on the sky in $x_s,z_s$ (pc) are plotted as blue ellipses with a jet precession model shown by a solid black curve. Model parameters are listed in Table \ref{tab:pmodel}. Here, celestial north points up and east to the left. 
    The yellow star on the bottom right marks the jet source as in Figure \ref{fig:arrow_image} and is placed at the origin. Molecular Hydrogen knots from \citep{Walther_2019} are approximated by a dashed curve, and the low-velocity CO cavity from \citep{Cheng_2022} is shown with gray arrows. The molecular hydrogen may mark a helix of gas accelerated by the jet, but we lack data to confirm this. The low-velocity CO gases encompasses a high-velocity CO jet and a radio continuum jet \citep{Trinidad_09, CarrascoGonzalez_2012} that respectively align with the molecular outflows and the beginning of our track of [Fe II] knots.
    \\
    \underline{\textit{Right}:} The $y_s$-coordinates for knots predicted from our model is shown with a dotted curve as a function of distance on the sky from the source, $d_{sky}$.  The $y_s$ coordinate gives distance along the line of sight with origin at the source and with positive $y_s$ more distant from the viewer.   
    \label{fig:modxy} }
\end{figure*}

We could not find a model that results in the observed deceleration (see Section \ref{outflowspeed}) via projection effects alone. Models with an extreme opening angle, wider than what we measure in Figure \ref{fig:arrow_image} (33$^\circ$), can exhibit a drop in proper motion as a function of distance from the source protostar. However, if our observed deceleration was caused by knots moving away from the observer, then model knot positions predicted on the sky tend to be closer to the source than observed. We find that precessing jet models only match knot locations and velocities when the ejected knots rapidly decelerate, as described with our exponential decay parameter  $\alpha$. Models that match the positions and proper motions have half opening angle of $\beta \approx 15^\circ$ and require moderate $\alpha$ values within about 20\% of that given in the Table \ref{tab:pmodel}. Such a model is shown in Figures \ref{fig:speed_trends} and \ref{fig:modxy} and has parameters listed in Table \ref{tab:pmodel}.

\begin{deluxetable}{lcc}
    \tablewidth{0pt}
    \caption{Precessing Jet Model Parameters \label{tab:pmodel}}
    \tablehead{ 
            \colhead{Parameter} & \colhead{Symbol} & \colhead{Fiducial Values}
        }
    \startdata
        Jet ejection velocity  &  $v_j$  & -325 km/s \\
        Precession period    &  $2\pi/|\Omega_p|$   & 2066 yr \\
        Jet decay rate & $\alpha$   & 1560 Myr$^{-1}$ \\
        Half jet opening angle & $\beta$  & 14.5$^\circ$ \\
        Binary orbit inclination & $i_B$   & 0$^\circ$ \\
        Position angle of precession axis & $\xi$   & 23.5$^\circ$ + 180$^\circ$ \\
        Initial phase & $\phi_0$   & 12.75$^\circ$ \\
        Mass of jet source & $m_1$   & 1 \Msun \\
        Mass of binary companion & $m_2$  & 1 \Msun\\
        Binary semi-major axis & $a_B$   & 20 au \\
    \enddata
\end{deluxetable}

Figure \ref{fig:modxy} shows knot positions on the sky from Table \ref{tab:knot_measures} as blue ellipses in the left panel. The black curve represents our model jet trajectory from HOPS 361-C. The plot axes $x_s$ and $z_s$ give coordinates on the sky in pc with origin at the jet source, which is shown with a fuchsia star. Celestial north is aligned with $z_s$, so it is facing up on this panel. Knot location can be used to estimate the time a knot was ejected from the source. Ejection times, estimated from this model, are shown on the top axis of Figure \ref{fig:speed_trends}.

Molecular Hydrogen knots from \citet{Walther_2019} are shown by an approximate dashed curve, and the low-velocity CO cavity from \citet{Cheng_2022} is shown with gray arrows. The helix of molecular hydrogen may be accelerated by the jet that we detect but cannot presently be analyzed in detail because of imaging differences (e.g. other protostars could heat gas, differences in brightness and epochs). We caution that past observations strictly correspond to any of the densest and warmest parts of the gas. The low-velocity CO gas encapsulates high-velocity CO gas and a jet of radio continuum emission from \citet{Trinidad_09, CarrascoGonzalez_2012} aligning with the molecular Hydrogen knots and lying on top of the track of our detected [Fe II] knots.

The right panel in Figure \ref{fig:modxy} shows the $y_s$ coordinate of the model jet trajectory as a function of distance on the sky from the source  ($d_{sky}$) as in Figure \ref{fig:speed_trends}. Along the line of sight, increasing $y_s$ corresponds to greater depths into the cloud.   
With our model, the jet is moving toward then away from the observer as a function of $d_{sky}$. 

HOPS 361-C was only recently resolved as a close binary system in Very Large Array (VLA) 9 mm continuum images with a model binary separation of about 0.1 arcsec or 43 au \citep{Cheng_2022}. Following \citet{Terquem_1999}, the binary system's orbital period is related to the jet's precessional period, assuming a rigid disk of gas within the binary system that precesses, uniform disk surface density, and Keplerian orbits,
\begin{equation}
    {\tau}_{orbit} = {\tau}_{prec} \frac{15}{32} \frac{q}{{(1+q)}^{1/2}} {\sigma}^{3/2} \cos(\beta),  \label{eqn:tau}
\end{equation}
where ${\tau}_{orbit}$ is the binary orbital period, $q=m_2/m_1$ is the ratio of the secondary mass divided by primary which is assumed to be the jet source. Here $\sigma$ is the ratio of $m_1$'s disk radius to the  binary orbital semi-major axis. Typically, $\sigma$ is taken to be 1/3, assuming tidal truncation, though it could range from 1/4 to 1/2.

We found precession models match positions and proper motion data with periods of $2\pi/|\Omega_p|$ ranging from about 1500--2500 yr. With a 2000 year precession period, a binary star comprised of two 1 \Msun \ stars ($q=1$) has an orbital period of 130 years and a semi-major axis of 33 au,  which is approximately consistent with the separation inferred from the VLA 9 mm images. However, kinematic modeling of molecular line emission gave an inner disk radius that is smaller, 10 to 25 au \citep{Cheng_2022}. If this inner radius encompasses the binary, then the binary semi-major axis would necessarily have to be smaller than the inner disk radius.

We note the precession model shown in Figures \ref{fig:speed_trends} and \ref{fig:modxy} is not sensitive to the binary separation or the mass of its constituents, though the small amplitude wiggles in the model curves are due to the binary orbital motion. With a larger binary separation, these wiggles would have lower amplitude and longer wavelength. 

Via kinematic modeling, \citet{Cheng_2022} estimate a binary mass for HOPS 361-C of $m_1 + m_2 \sim 1.5 M_\odot$, though modeling the spectral energy distribution (SED) gave a protostellar mass of $\sim 2 M_\odot$ neglecting binarity. These estimates for the protostellar masses are similar to our assumed values of $m_1=m_2$ and $m_1 + m_2 = 2 M_\odot$. Based on an estimate for the location of the kinematic center, \citet{Cheng_2022} estimate a mass ratio $m_1/m_2 \approx 1.3$ --2.5 which is within a factor of 2 of our assumed value of $m_1/m_2 = 1$. Using our current measurements and uncertainties, the jet precession model is not inconsistent with what is known about the HOPS 361-C binary system.

The direction of precession is predicted to be in the opposite direction as the rotation of the disk from which the jet originates ($\Omega_p$ is negative in equation 1 by \citealt{Terquem_1999}). Since the precession rate is computed by averaging the tidal force over the binary's orbit, it is independent of the direction of rotation in the binary orbit. We take $\Omega_p <0$ so that position angle and inclination are given with respect to the binary orbit normal.  We assume the direction of rotation for the disk driving the jet is similar to the direction of rotation of the binary orbit (for which we chose $\beta <\pi/2$). The position angle of the precession axis $\xi$, is also the position angle of the binary orbit normal. This axis points to the south, consistent with the direction of rotation seen in the disk resolved by \citet{Cheng_2022}. 

The CO J = 2–1 data for HOPS 361-C also shows a high-velocity, bipolar jet with a position angle of $22$ to $32^\circ$. This differs from the radio jet axis seen at 9 mm that has a position angle of $\sim 15^\circ$ \citet{Cheng_2022}. Near the source,  our precession model has the jet oriented near its maximum opening angle, giving a jet position angle of about $10^\circ$ and approximately consistent with the radio jet axis.

The jet opening angle $\beta$ is sensitive to the solid angle extended by the arc with respect to our origin, the protostellar source. The model is less sensitive to the binary inclination $i_B$, but we found models that matched both proper motions and knot positions with $i_B< 25^\circ$. Radial velocity from the [O I] spectrum restricted $i_B$ to a few degrees. \citet{Cheng_2022} fit kinematic models to their ALMA molecular line data for the HOPS 361-C disk-jet system, and they estimated a disk inclination of 63$^\circ$ to 77$^\circ$ and a position angle of around 15$^\circ$ for the radio jet. If we assume their disk inclination is that of a circumbinary disk in a plane containing the binary, this inclination corresponds to $i_B \sim 90-70 = 20^\circ$ for the inclination of the binary orbit normal vector using our variable which has zero inclination for an edge-on binary orbit.  Our precession model is consistent with the estimates for the disk inclination by \citet{Cheng_2022}, assuming their model pertains to a circumbinary disk.  

The disk resolved by \cite{Cheng_2022} has a blue shifted component on the western side and red-shifted component on the eastern side, which suggests that the binary orbit's angular momentum vector is pointed to the south. To compare, if the binary orbit and the disk that drives the jet have the same direction of rotation (opening angle $\beta < \pi/2$), then the jet's arc is part of a counter jet that propagates in the direction opposite to the binary orbit's normal. The position of the counter jet can be predicted using equation \ref{eqn:vjte} with $v_j<0$, giving the expected S-shaped symmetry about the origin between jet and counter jet.  This is why we have chosen negative $v_j$ for our model in Table \ref{tab:pmodel}. 

The direction of precession in our model is consistent with a disk-driven jet having rotation direction similar to the presumed circumbinary disk observed in various molecular lines by \citet{Cheng_2022}. We concur with \citet{Cheng_2022}, who concluded that jet precession and associated interactions between jet and environmental material at different times are the most likely scenario accounting for the observed differences in the jet position angles seen with different tracers.

   \label{precess}

   \subsection{Outflow Depth in Surrounding Environment}

We achieve a quasi-3D positioning of the knots by comparing foreground extinction to each knot in Figure \ref{fig:speed_trends} (or see Table \ref{tab:knot_measures}) with the extinction to material at the back of the molecular cloud. Their ratio gives an approximate value for how deeply embedded each knot is within the cloud. But relative distances may change due to radial distance, column density, number density, or opacity. Therefore, they may not directly correspond to the radial positions predicted by our precession model (the right panel of Figure \ref{fig:modxy}).

Extinction to the back of the cloud can be determined using molecular gas, like optically thin CO isotopologues, that traces relatively slow motions  \citep[e.g.][]{Dickman_1975, Dickman_1978, Goldsmith_1992, Schwartz_1983, Gutermuth_2008, Stojimirovic_2008}. 
We use the correlation between extinction and $\rm {C}^{18}$O (J = 1--0) molecular line emission \citep{Alves_1999}
\begin{equation}
    A_V = \frac{I({\rm C}^{18}{\rm O}) + 0.4\pm0.1}{0.21\pm0.01}.  \label{eqn:AVCO}
\end{equation}
The region of the integrated $\rm {C}^{18}O$ map from \citet{Iwata_1988} that overlaps our NIR images 
gives a brightness temperature of 2.52--2.8 $\rm K \ km \ {s}^{-1}$, which corresponds to an \Av \ of 13.9--15.2 mag using equation \ref{eqn:AVCO}. We only find a single value for the region since the radio observations have a nearly 1 arcmin spatial resolution, nearly the scale of our jet. We could use $\rm {C}^{18}O$ data in \citet{Schwartz_1983}, but their column density maps have fewer contour lines to use for measurements.

We present the fraction of each knot's \Av\ relative to total \Av\ through the cloud in the bottom panel of Figure \ref{fig:speed_trends}. Translating the contours from \citet{Iwata_1988} to an \Av \ value with the correlation from \citet{Alves_1999} introduces an additional uncertainty of approximately $\pm$1.6 mag to the uncertainties in the extinction, and we have used this uncertainty to estimate error bars for the points plotted in the bottom panel of Figure \ref{fig:speed_trends}. 

Fractional depths in the bottom panel of Figure \ref{fig:speed_trends} indicate knots oscillate between being more and less embedded, potentially moving closer and further from the viewer. We take caution making firm constraints with our precession model, since the cloud is unlikely to be perfectly uniform \citep[e.g.,][]{Schwartz_1983}. But precession may impact jet morphology and dynamics more than variations in a jet's gas density \citep[e.g.][]{CastellanosRamirez_2018}. Quantitatively, the knots are at a fractional depth of 1/5 to 4/5 into the molecular cloud, where they may disrupt the HOPS 361 region.
    \label{compare_av}  

    \subsection{Jet Mass, Momentum, and Energy Injection}
We search for effects that connect jet properties relevant to stellar feedback, our precession model values, and HOPS 361 cloud clump properties. 
Assuming a 1D, incompressible fluid flow and mass conservation across each knot (ignoring changes in viscosity, plasma density, and mass flow rate) as in the MAPPINGS model, the jet has 
%
%
a mass outflow rate or jet output (\Mdot) of
\begin{equation}
    {\dot{M}} = \frac{\Delta M}{\Delta t} = {\rho}_{j} { A} {{v}_{outflow}},
\label{eqn:Mdot}
\end{equation}
%
momentum flux (\Pdot) or ram pressure through each knot for constant \Mdot \ as
\begin{equation}
    {\Delta \dot{P}} = \frac{\dot{M}}{A} \Delta v = \frac{\dot{M}}{A} {({v}_{outflow} - {v}_{P})} = \frac{\dot{M}}{A} {{v}_{S}}, 
    \label{eqn:Pdot}
\end{equation}
 and a flux of kinetic energy ($\dot K$) or energy density dissipated in the shock in  each knot with constant \Mdot \ of
\begin{equation}
    {\Delta \dot{K}} = \frac{1}{2}~\frac{\dot{M}}{A} \Delta(v^2) 
    = \frac{1}{2}~\frac{\dot{M}}{A} {({{v}_{outflow}}^2 - {{v}_{P}}^2)} 
    \label{eqn:KEdot}
\end{equation} 
Here 
${\rho}_{jet}$ is the jet's mass density, ${{v}_{outflow}}$ is total flow speed through each knot, ${v}_{P}$ is knot proper motion, ${v}_{S}$ is shock speed, and $A$ is the jet's cross-sectional area. We compute $A = \pi {{R}_{knot}}^2$, taking ${{R}_{knot}}$ to be the semi-minor axis from Table \ref{tab:knot_measures} for an ellipse encircling each knot. 
As in section \ref{jetmod}, we estimate $v_{outflow}$ from the sum of the knot proper motion and shock speed. 
Assuming the majority of the jet's mass consists of neutral atomic hydrogen, we find the mass density by multiplying the jet's number density (${n}_{0} \approx 3.2 \times 10^4\ {\rm cm}^{-3}$ or see Section \ref{jetmod}) by the mass of a hydrogen atom and using a mean molecular weight of 1. 
We estimate the total mass density in the jet is ${\rho}_{jet} = 5.34\times {10}^{-20} \ {\rm g} \ {\rm cm}^{-3} = 789$ \Msun \ ${\rm pc}^{-3}$.
We list each knot's \Mdot, \Pdot, and $\dot{K}$ \ multiplied by area $A$ for ease of comparison \ in Table \ref{tab:feedbackvals}.

\begin{deluxetable*}{ccccc}
    \tablewidth{0pt}
    \tablecaption{Derived Feedback Properties for Each Knot}
    \tablehead{
     \colhead{Knot Identifier} & \colhead{$A$} & \colhead{\Mdot}  & \colhead{$\Delta$\Pdot $\times A$}  & \colhead{$\Delta \dot K \times A$}  \\
     \colhead{...} & 
     \colhead{$\rm {pc}^{2}$} &
      \colhead{\Msun \, $\rm {yr}^{-1}$}  &
      \colhead{\Msun \, $\rm {yr}^{-1}$ km $\rm {s}^{-1}$}  &  
      \colhead{\Lsun}   
    }
    \startdata
        1	&   7.69$\times{10}^{-6}$   &   2.12$\times{10}^{-6}$   &	1.10$\times{10}^{-4}$	&   5.69 \\
        2	&   6.62$\times{10}^{-7}$   &   1.32$\times{10}^{-7}$   &	6.61$\times{10}^{-6}$	&   0.240 \\
        3	&   1.29$\times{10}^{-5}$   &   2.93$\times{10}^{-6}$   &	1.38$\times{10}^{-4}$	&   5.85 \\
        4	&   2.77$\times{10}^{-6}$   &   4.29$\times{10}^{-7}$   &	2.19$\times{10}^{-5}$	&   0.598 \\
        5	&   3.28$\times{10}^{-6}$   &   4.62$\times{10}^{-7}$   &	2.27$\times{10}^{-5}$	&   0.558 \\
        6	&   6.90$\times{10}^{-6}$   &   9.72$\times{10}^{-7}$   &	5.08$\times{10}^{-5}$	&   1.24 \\
        7	&   5.78$\times{10}^{-6}$   &   7.56$\times{10}^{-7}$   &	3.97$\times{10}^{-5}$	&   0.885 \\
        8	&   4.29$\times{10}^{-6}$   &   4.62$\times{10}^{-7}$   &	2.21$\times{10}^{-5}$	&   0.397 \\
        9	&   1.16$\times{10}^{-5}$   &   1.33$\times{10}^{-6}$   &	6.57$\times{10}^{-5}$	&   1.27 \\
        10	&   2.17$\times{10}^{-5}$   &   1.79$\times{10}^{-6}$   &	8.91$\times{10}^{-5}$	&   1.13 \\
        11	&   1.54$\times{10}^{-5}$   &   1.24$\times{10}^{-6}$   &	5.99$\times{10}^{-5}$	&   0.740 \\
    \enddata
    \tablecomments{Jet properties used to evaluate how the HOPS 361-C jet affects its host cloud.  The cross-sectional area derives from squaring the smaller, minor axis of the ellipse centered on each knot and using values from Table \ref{tab:knot_measures}. The flow speeds needed for other columns are listed in Table \ref{tab:propmot}.
     \label{tab:feedbackvals}}
     \vspace{-0.2in}
\end{deluxetable*}

According to \citet{Watson_2016} and \citet{Sperling_2021}, the jet output efficiency is the ratio of the mean mass outflow rate through the jet to the accretion rate. 
Mass accretion rate can be estimated for Class 0 and Class I protostars using bolometric luminosity (${L}_{bol}$) and stellar properties (mass, ${M}_{\ast}$, and luminosity, ${L}_{\ast}$) by assuming that the bolometric luminosity is equal to the total luminosity for the binary system as well as the accretion luminosity, then
\begin{equation}
    {\dot{M}}_{acc} = \eta \frac{{L}_{bol} {R}_{\ast}}{G {M}_{\ast}}
\end{equation}
\citep{Enoch_2009, Evans_2009}, where $\eta$ is an efficiency factor often assumed to be on the order of 1 \citep{Fischer_2017, Muzerolle_2003, Calvet_1998, Meyer_1997}. 
Using the methods from \citet{Furlan_2016} to estimate bolometric luminosity, 
\citet{Cheng_2022} estimated a mass accretion rate of $9 \times {10}^{-6}$ \Msun \ $\rm {yr}^{-1}$ for HOPS 361-C.   
Using, the mean \Mdot\ from Table \ref{tab:feedbackvals} of 1.15 $\times {10}^{-6}$ \Msun \ $\rm {yr}^{-1}$ and the above accretion rate estimate, we find an efficiency of 0.128.  
Efficiency is related to the location of the jet launching site, with larger efficiencies corresponding to launch at a smaller radius \citep{Watson_2016, Sperling_2021}.
Based on \citet{Watson_2016} and \citet{Sperling_2021}, values around 0.1 cannot rule out any jet launching models. If the bolometric luminosity is split between the components of the binary system, then the efficiency may increase, and jet launching models that place the footpoint closer to the protostar may better explain why our value exceeds 0.1. 
 
We determine the momentum injection rate 
through the jet imparted into the surrounding cloud, compare with escape velocity, and find the amount of mass the jet ejects from the HOPS 361 clump \citep[as in][]{Matzner_2007, Arce_2010, Li_2020}. We sum $\dot{P} \times A$ values for each knot listed in Table \ref{tab:feedbackvals}. Since the interaction with the surroundings may not be perfectly efficient or uniform, then the total force is $\leq$6.26 $\times {10}^{-4}$ \Msun \, $\rm {yr}^{-1}$ km $\rm {s}^{-1}$. 
Taking the clump as a uniform, non-rotating, non-magnetic, spherical mass of $M_c=88.9$ \Msun \ at a radius of $R_c=0.2$ pc (using the Herschel map of Orion B described in \citealt{Stutz_2016}, Stutz p. com., and \citealt{Furlan_2016}, and the correlation between N(H) and extinction found by \citealt{Pillitteri_2013}), then the clump's escape velocity ($v_{esc}$) is 2 $\rm km \ s^{-1}$. The ratio of the sum of momenta from the jet's knots to this escape velocity gives \Mdot$_{esc}$ for material escaping the cloud equal to 3 $\times 10^{-4}$ \Msun \ $\rm yr^{-1}$, assuming constant escape velocity and momentum output. Over a typical protostellar lifetime or a freefall timescale of 100,000 years, then the jet would launch approximately 30 \Msun \ out of the system, or about a third of the clump's mass. Therefore, a few precessing jets from intermediate-mass protostars may have enough momenta to eject the majority of the clump's mass at this scale.
To find the effect of the jet's kinetic energy, we sum power (${\dot K} \times A$) emitted by each knot in Table \ref{tab:feedbackvals} to get a total of 18.6 \Lsun. 
Gravitational potential energy an extended mass or cloud, $M_c$, with radius, $R_c$, is generally
\begin{equation}
    U = -C \frac{G {M_c}^2}{R_c},
    \label{eqn:U}
\end{equation}
where $C$ is a constant dependent on the cloud's geometry and uniformity. Assuming the mass and radius to compute our escape velocity, equation \ref{eqn:U} gives $1.21 \times {10}^{45} \, \rm erg$. If the cloud is virialized and only has a total of half of this energy, then the ratio of half of $U$ to ${\dot K} \times A$ is a 480 yr timescale. This timescale may suggest a precessing jet contributes a significant portion of a cloud clump's gravitational energy budget during a protostellar lifetime. We caution this is the fastest possible case with our present assumptions (perfectly efficient energy transfer, uniform cloud, constant output). 


The jet's power may contribute to turbulence within the molecular cloud. The dissipation rate due to turbulence at a velocity of $v_{turb}$ in the molecular cloud is 
\begin{equation}
    L_{turb} \sim \frac{ M_{c} v_{turb}^3}{l_{eddy}},
\end{equation}
where $l_{eddy} $ is the size of the largest eddies within the cloud \citep[e.g.][]{Stone_1998, Elmegreen_2004, Quillen_2005}. 
Taking the NGC 2071 cloud mass $M_{c} \sim 2000$ \Msun and $l_{eddy} \sim 1 $ pc (from ${}^{\rm 12}$CO and $\rm {}^{13}$CO maps; see \citealt{Stojimirovic_2008})
, $v_{turb} \sim 1.9$ km/s
(based on $\rm C^{18}O$ line widths by \citealt{Iwata_1988} and \citealt{Stanke_2022}; also a velocity FWHM of 1 km/s from an NH$_3$ line map using the Green Bank Telescope, by priv. comm. with J. Di Francesco), then 
we estimate a turbulent dissipation luminosity of about $2.3$ \Lsun.
Since the knots deliver an average power of about 1.69 \Lsun\ into the cloud,
the energy dissipation rate via turbulence in the cloud is similar to the average power dissipated locally by the jet. The jet itself may have enough energy to locally drive molecular cloud turbulence.

The HOPS 361-C jet may have enough momentum and energy to disrupt the HOPS 361 protostar-forming clump, but that depends on how efficiently the jet outputs material over time. The jet dissipates over a distance of $\sim$0.2 pc (Figures \ref{fig:speed_trends} and \ref{fig:modxy}). A knot launched at 325 km/s traveling 0.2 pc is a dynamical timescale of 600 years, which is similar to our damping timescale, the reciprocal of $\alpha$, of 640 yr. 
The damping or dynamical timescale for knots can occur about 3 times within our precession timescale of 2000 years, so the jet may extend further as seen in Figure \ref{fig:mosaics_epoch2}. However, knots outside this 0.2 pc distance may be due to other outflows or may be associated with lower velocity shocks $<$20 km/s, so future observations are needed to determine their source. Our study of the arced jet associated with HOPS 361-C suggests precessing jets with wide opening angles are locally damped within 0.2 pc of their driving source.



    \label{feedback} 
    
    \subsection{Criterion Allowing a Protostellar Jet to Puncture a Molecular Cloud}

What condition allows subsequently emitted clumps in a jet to dissipate or remain within the cavity opened by previously emitted clumps? 
For small opening angles, clumps in protostellar outflows lie on a similar path as the jet itself, and these jets open a cavity within a molecular cloud \citep{Quillen_2005, Cunningham_2009b,Frank_2014,Fendt_2022}. Long jets with small wiggles can puncture their host molecular cloud \citep[e.g. simulations by ][]{Velazquez_2013} and extend more than 1 pc from their source \citep{Bally_1982,Bally_2016}. In contrast, instead of passing  through the molecular cloud, the HOPS 361-C jet seems to entirely dissipate locally within it. 

We estimate the speed that a cavity is opened by a jet and compare it to its advance speed. 
The radial distance and radial velocity from the jet's central axis of symmetry  are
\begin{align}
r(t) &= v_j \sin(\beta) t \\
\dot r(t) &= v_j \sin(\beta) \sim \rm constant,
\label{eqn:dotr}
\end{align}
where $v_j$ is the jet speed and $\beta$ is the half opening angle.
A jet cavity that would not fill in at the speed of sound, $c_s$, in the ambient molecular cloud would satisfy
\begin{equation}
    \dot r \lesssim c_s.
\end{equation}
This corresponds to a jet cone that moves perpendicular to its central axis slower than the sound speed.
Combining this equation with equation \ref{eqn:dotr}, a jet cavity that does not fill in would satisfy
\begin{equation}
    \sin(\beta) \lesssim \frac{c_s}{v_j}. \label{eqn:constraint}
\end{equation}
Using a jet velocity of $v_j = 400$ km/s and 
a sound speed for the cloud's interstellar medium in the of $c_s \sim 1$ km/s, we estimate the half opening angle $\beta \lesssim 1^\circ$ for a jet that can maintain a cavity. 
In this scenario, any jet traveling at a few 100 km/s within an opening angle of a degree requires motions above the local sound speed to fill in the associated cavity. If the opening angle is larger than a degree, then the cavity would fill in.  Opening a cavity requires energy.  A jet with a wide opening angle may then expend more energy to propagate. This suggests wider jets would dissipate more rapidly, and long jets with large opening angles should be rare. 

This scenario ignores turbulence and magnetic fields.  But note that the supersonic, turbulent speed of gas in the interstellar medium is approximately Mach 3, where the Mach number is the ratio of  motion in the cloud to the local sound speed \citep{Ballesteros_2007}.
To take into account turbulence in the molecular cloud, we multiply the right hand side of equation \ref{eqn:constraint} by a factor of 3, and this would relax the restriction on the opening angle for a jet that can puncture a molecular cloud. 

2D magneto-hydrodynamic simulations have investigated how a precessing jet nozzle affects the propagation of a high-speed jet \citep{Fendt_2022}. Jets with wide opening angles ($>20^\circ$) fail to propagate to large distances as they are dissipated locally (see their Figure 6) but suggest the condition allowing a jet to propagate is less restrictive that by equation \ref{eqn:constraint}. Future simulations could relate protostellar jet observations (e.g. knot proper motions) to other cloud, jet, and binary properties and elucidate connections between jet opening angle and jet dissipation. Simultaneously observing molecular, atomic, and ionized components of the HOPS 361-C jet, including potential followup for proper motions, may also confirm $\rm H_2$ knots accelerated and carved out by HOPS 361-C.

    \label{fillfactor} 
    
\section{Conclusion} 
    We study new narrowband HST images of the NGC 2071 IR/HOPS 361 star-forming region. To summarize results proper motions of 350--100 km/s suggest the arc of knots bright in [Fe II], which rapidly decelerate away from NGC 2071 IR, appear to trace back to the protostar HOPS 361-C. We measured a radial velocity of 3 km/s using a new SOFIA [O I] 63 \mic \ spectrum near the jet’s base and found this jet is nearly in the plane of the sky. Knots with proper motions are also confirmed to be shocked with typical speeds of 50 km/s and densities of 3 $\times \ {10}^4 \ {\rm cm}^{-3}$, and from extinctions they are embedded at depths of 1/5 to 4/5 into the cloud. 

We apply the precession model by \citet{Masciadri_2002} to infer the full kinematics of the jet from the knot positions, knot speeds, and measured opening angle of 16$^\circ$. We need an additional parameter to describe the jet's deceleration, assuming the jet has a constant ejection velocity. Our model supports prior proposals that this jet is precessing with a period of 2000 yrs, and this is consistent with binary system properties estimated for HOPS 361-C \citep{Cheng_2022}. A knot at 0.13 pc from HOPS 361-C may have been deflected, launched by another protostar, or be a spurious motion. 

Instead of precession due to the secondary in a binary system, an arc of knots can alternatively be produced by asymmetric infall of the envelope onto the protostar \citep{Hirano_2019, Lee_2020} or tidal encounters \citep{Cunningham_2009b}. Observing the binary source and its envelope could help differentiate asymmetric infall, and future radial velocity measurements for more distant knots confirming or rejecting our model could test whether another option to induce precession is more viable. Another epoch of observations for the wider field for knots more distant than 0.2 pc from the source (on the sky) would enable measuring proper motions, which would show whether the proper motions have dissipated rapidly, or perhaps if more distant knots are associated with outflows from a different protostar.
  
The HOPS 361-C jet illustrates how jet precession can affect stellar feedback and the host cloud. Lower shock velocities past the end of the arc and more than 0.2 pc from HOPS 361-C suggest that the jet may have almost entirely dissipated at this distance. This is consistent with local and rapid decay of the jet's kinetic energy except if the jet's ejection velocity increases with time. Prior studies have investigated protostellar outflow-induced feedback into molecular clouds \citep[e.g.][]{Matzner_2007, Raga_2009, Carroll_2009, Federrath_2014,Frank_2014, Nakamura_2014, Rohde_2022}. The short 0.2 pc dissipation length in HOPS 361-C's jet and total kinetic energy suggests that precessing binary systems could sustain and affect the distribution of length scales for outflow-induced turbulent energy injected into NGC 2071 IR. 

Since many known jets with constant direction or small opening angles extend much further than 0.2 pc, and do not decrease as quickly in velocity as a function of distance from their source \citep[e.g., ][]{Lee_2020,Erkal_2021}, we suspect that the rapid deceleration we see in the HOPS 361-C jet arc is associated with the jet's wide opening angle. If so, the momentum and kinetic energy in a wide opening angle precessing jet can be imparted to the molecular cloud closer to the jet source than that of a narrow opening angle jet. Future models, building on simulations of precessing jets \citep[e.g.][]{Fendt_2022}, could explore how binary stars that produce precessing jets at different outflow rates impact feedback into molecular clouds.

\section{Acknowledgments} 
We thank Eric Blackman and Jonathan Carroll-Nellenback for discussions. We are grateful to Liam Hainsworth for calculating MAPPINGS model grids. Finally, we thank our referee's insights and perspectives.

This research is based on observations made with the NASA/ESA \textit{Hubble Space Telescope} obtained from the Space Telescope Science Institute (STSci), which is operated by the Association of Universities for Research in Astronomy, Inc., under NASA contract NAS 5–26555. These observations are associated with programs 11548 and 16493. AER, NK, and SF were supported by funding from STScI from program 16493. 

The data in this paper are obtained from the Mikulski Archive for Space Telescopes (MAST) at the STSci. The specific observations analyzed can be accessed via \dataset[10.17909/e0g3-wc67]{https://doi.org/10.17909/e0g3-wc67} \& \dataset[10.17909/fwz1-0q60]{https://doi.org/10.17909/fwz1-0q60}.


This work is based in part on observations with the NASA/DLR SOFIA. SOFIA is jointly operated by the Universities Space Research Association, Inc. (USRA), under NASA contract NNA17BF53C, and the Deutsches SOFIA Institut (DSI) under DLR contract 50 OK 2002 to the University of Stuttgart. 

This work uses \textit{Montage}, which is funded by the National Science Foundation (Grant Number ACI-1440620). \textit{Montage} was previously funded by the NASA Earth Science Technology Office, Computation Technologies Project, under Cooperative Agreement Number NCC5-626 between NASA and Caltech.



    
\section{Appendix}\label{sect: Appendix}
    \begin{figure*}
        \centering   
        \includegraphics[width=\textwidth, trim = 0in 0in 0in 0in,clip]{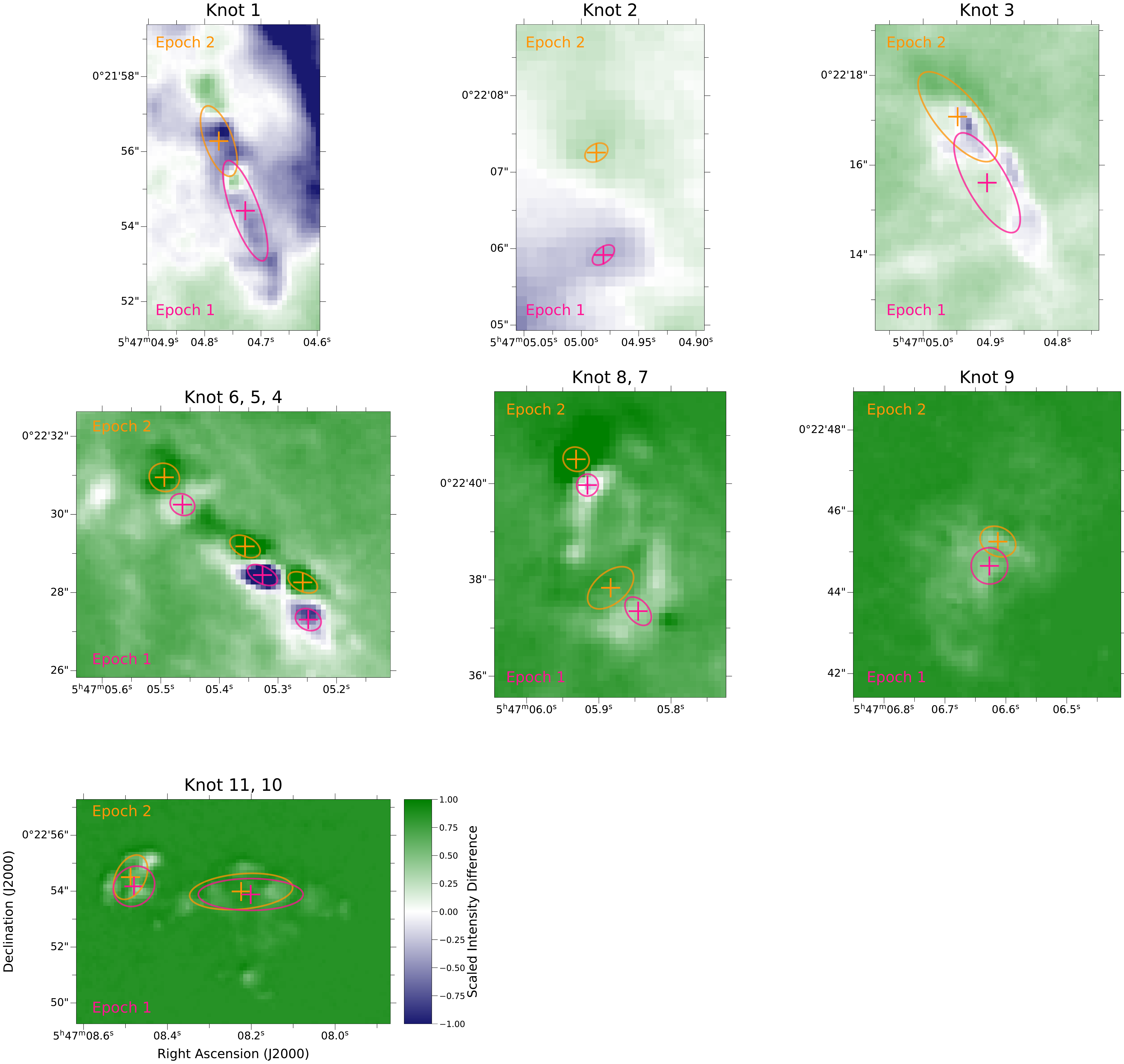}
        \caption{
        Close-ups of knots identified from the difference image in Figure \ref{fig:compare_epochs}. Epoch 1 is shown in pink and Epoch 2 is shown in orange. Ellipses show the regions associated with each epoch and knot. Pairs of + marks are the positions from Table \ref{tab:obs}, which generally move up or to the left. The direction of motion and pixel shifts in Table \ref{tab:propmot} are measured with a line connecting the + marks for each knot.
        }\label{fig:app_zoomin}
    \end{figure*}
    \clearpage
    
\bibliography{biblio}{}
\bibliographystyle{aasjournal}
    
\end{document}